%% file: main.tex
\documentclass[apj, floatfix,table, twocolumn]{aastex631} %twocolumn
\usepackage{amsmath}
\usepackage{xspace}
\usepackage{CJKutf8}
\usepackage{ulem}
\usepackage[]{xcolor}
\definecolor{meridithgreen}{RGB}{0, 150, 0}

\newcommand{\CNnames}[1]{{\begin{CJK}{UTF8}{gbsn}~(#1)~\end{CJK}}}

\newcommand{\mmode}[1]{\ifmmode{#1}\else{$#1$}\fi}
\newcommand{\Dnu}[0]{\mmode{\Delta\nu}}
\newcommand{\DPi}[1]{\mmode{\Delta\Pi_{#1}}}

\newcommand{\fDnu}[1]{\mmode{f_{\Delta\nu}}}
\newcommand{\numax}[0]{\mmode{\nu_\text{max}}}
\newcommand{\Teff}[0]{\mmode{T_\text{eff}}}
\newcommand{\logg}[0]{\mmode{\log g}}

\newcommand{\Msun}[0]{\mmode{\text{M}_{\odot}}}
\newcommand{\Lsun}[0]{\mmode{\text{L}_{\odot}}}

\newcommand{\dm}[0]{\mmode{\varepsilon_{\texttt{dmesh}}}}
\newcommand{\dPm}[0]{\mmode{\zeta_{\texttt{dmesh}}}}
\newcommand{\dmesh}[0]{\mmode{\texttt{delta\_mesh\_coeff}}}
\newcommand{\Kepler}[0]{Kepler}

\newcommand{\TESS}[0]{TESS}

\newcommand{\Aosc}[0]{\mmode{A_{\rm osc}}}

\begin{document}

\title{Beyond \texttt{MESA} Defaults:\\The Impact of Structural Resolution Uncertainty in $p$-mode Asteroseismology} 
% Beyond MESA Defaults: The Impact of Structural Resolution Uncertainty in p-mode Asteroseismology

\author[0000-0003-3020-4437]{Yaguang~Li\CNnames{李亚光}$^{\star}$} 
\email{yaguangl@hawaii.edu}
\affiliation{Institute for Astronomy, University of Hawai`i, 2680 Woodlawn Drive, Honolulu, HI 96822, USA}

\author[0000-0002-8717-127X]{Meridith Joyce$^{\star}$}
\email{mjoyce8@uwyo.edu}
\affiliation{University of Wyoming, 1000 E University Ave, Laramie, WY USA}
\affiliation{Konkoly Observatory, HUN-REN CSFK, Konkoly-Thege Mikl\'os \'ut 15-17, H-1121, Budapest, Hungary}
\affiliation{CSFK, MTA Centre of Excellence, Konkoly-Thege Mikl\'os \'ut 15-17, H-1121, Budapest, Hungary}

\begingroup\def\thefootnote{$\star$}
\footnote{These authors contributed equally to this manuscript.}
\endgroup

\begin{abstract}
Observations of pressure modes ($p$-modes) in stars have enabled profound insights into stellar properties, and theoretical stellar evolution and oscillation models are integral to these inferences. However, modeling uncertainties are often overlooked, even as they can rival or exceed observational uncertainties. In this study, we quantify, for the first time, the impact of structural resolution choices in 1D stellar evolution calculations on predicted $p$-mode frequencies across the HR diagram, using \texttt{MESA} and \texttt{GYRE}. 
We present demonstrative measurements of resolution-based modeling uncertainty for a range of solar-like, upper main-sequence, and Mira oscillators and compare these directly to \TESS{} observational uncertainties. 
We demonstrate that resolution-driven uncertainties can significantly influence theoretical predictions and in some cases overwhelm observational uncertainties by orders of magnitude. For the illustrative case considered---an order-of-magnitude variation in mesh resolution---solar-like oscillators typically have fractional, resolution-based uncertainties at or below 1\% of the test frequency. Fractional uncertainties in Miras, however, are as large as 20\%. 
We also find that the location and morphology of the RGB bump and red clump are impacted substantially by resolution uncertainty. 
Stellar ages are impacted at the 10\% level for young main-sequence stars, and the model-based correction factor for the $\Dnu{}$--$\sqrt{\rho}$ scaling relation is impacted at the 2\% level.
Our results underscore the need to incorporate modeling uncertainties into asteroseismic analyses and provide a reference framework for observers evaluating the reliability of theoretical models.
\end{abstract} 

\section{Introduction}
\label{sec:intro}
Pressure modes, or $p$-modes, have been observed in stars across a broad spectrum of masses, stellar types, and evolutionary phases. Such oscillations serve as a powerful tool for stellar characterization, enabling the inference of sub-surface structural information that can reveal the inner workings of stellar interiors \citep{Aerts2021} and more reliably constrain fundamental properties like mass and age 
\citep{Chaplin2013,SilvaAguirre2015,Kurtz2022}. Theoretical calculations from stellar evolution and stellar oscillation codes constitute a critical component of the data-to-parameter pipeline. However, such modeling instruments are rarely used with the same standards of consideration afforded to observational instruments. 

For some well-characterized stars \citep[e.g. $\alpha$ Centauri A \& B, 16 Cyg A \& B;][]{Kjeldsen2005-acen,deMeulenaer2010-acen,Metcalfe2012,Davies2015,Joyce2018-acen}, the precision on observed frequencies quoted for individual $p$-modes exceeds the reliability of frequencies predicted by stellar models by several orders of magnitude \citep[e.g.][]{Murphy2021, Cunha2021}. Despite the fact that modeling uncertainty overwhelmingly dominates the error budget in most cases concerning fits to individual $p$-modes, modeling uncertainties are rarely incorporated into asteroseismic results.

One reason modeling uncertainty considerations are commonly avoided is that the concept of ``modeling uncertainty'' encompasses at least three different issues:
\vspace{-0.1cm}
\begin{enumerate}
    \item predictions differ across codes; \vspace{-0.2cm}
    \item predictions differ when physical assumptions are varied within the same code; and \vspace{-0.2cm}
    \item predictions differ when numerics, such as spatial resolution, temporal resolution, or solution tolerance, are changed within the same code. 
\end{enumerate}
Several authors have examined the first of these, both in the context of fits to individual objects \citep{Huber2024} and to quantify variance across codes independent of any particular target \citep{Christensen-Dalsgaard2020}.
Other studies have focused on the second issue, often using observationally well-characterized objects to underscore the importance of the intrinsic uncertainty in our modeling \textit{choices} relative to observational precision 
(e.g.\ \citealt{Joyce2018-mp,Joyce2018-acen, Tayar2022, YingChaboyer2023, Joyce2023,Cinquegrana2023,Liyg2024}; L. M. Morales, J. Tayar \& Z. R. Clayton, in preparation). 
The third of these issues is addressed in the form of convergence studies, typically relegated to the appendices of our most robust modeling efforts. Resolution and convergence studies are not as common as they should be, but even when they are performed, the results are necessarily specific to the regime of the problem. While convergence tests may reinforce the reliability of results over a limited domain of interest, they do not tell us much about the overall reliability of software instruments. 

A complete characterization of modeling uncertainty would thus involve rigorously quantifying differences between tools, differences between physical assumptions, and differences between numerical choices. This is understandably perceived to be beyond the scope of most analyses, especially in the case of point 1, given that many software tools are closed-source (see \citealt{Tedersoo2021}).
However, the danger of neglecting modeling uncertainty in our asteroseismic analyses worsens as observational precision improves and as our data-driven inferences become more contrived and less transparent (e.g.\ using machine learning for bulk parameter estimation).

The Modules for Experiments in Stellar Astrophysics (\texttt{MESA}; \citealt{mesa2011,mesa2013,mesa2015,mesa2018,mesa2019,MESAVI}) software suite is widely used in conjunction with the \texttt{GYRE} stellar oscillation program \citep{GYRE} to compute asteroseismic properties for stellar models. Thanks to space-based photometric missions like WIRE \citep{Hacking1997}, MOST \citep{Matthews2000}, CoRoT \citep{Auvergne2009}, Kepler \citep{Borucki2010}, and TESS \citep{Ricker2015},
astronomers are obtaining increasingly high-precision measurements of individual $p$-mode pulsation frequencies for stars whose modes can be reliably identified. This data climate demands a more realistic understanding of the uncertainty in theoretical frequency predictions.
We therefore focus in this paper on one of many under-explored contributions to theoretical error budgets: structural resolution in 1D stellar evolution and oscillation calculations. 

In this study, we quantify the impact of varying \texttt{MESA}'s structural resolution control, \texttt{delta\_mesh\_coeff}, on theoretical $p$-mode frequencies across the HR diagram. We compare frequencies for solar-like oscillators (surface turbulence-driven), upper-main sequence pulsators ($\kappa$-mechanism-driven, including $\delta$ Scutis), and Miras ($\kappa$-mechanism-driven), covering evolutionary regimes from the main sequence to the asymptotic giant branch.

In Section \ref{sec:methods}, we discuss our model grid and physical assumptions, describe our method for comparing evolutionary tracks with different structural resolution, and introduce our method for measuring the discrepancies in frequency and period caused by resolution choices. 
In Section \ref{sec:results}, we present our resolution uncertainty calculations across the HR diagram. In Section \ref{sec:TESS}, we compare this source of modeling uncertainty directly to \textit{TESS} observational uncertainties for stars of the appropriate type. \textbf{It is intended that this section serve as a quick-look reference for observers interested in a first-order estimate of the reliability of models in the regime of their data.}
In Section \ref{sec:impacts}, we discuss how our uncertainty considerations would impact modern asteroseismic techniques. We conclude by making recommendations for best practices in future asteroseismic $p$-mode analyses in Section \ref{sec:best_practices}.

\section{Methods}
\label{sec:methods}
\begin{figure} 
\includegraphics[width=\columnwidth]{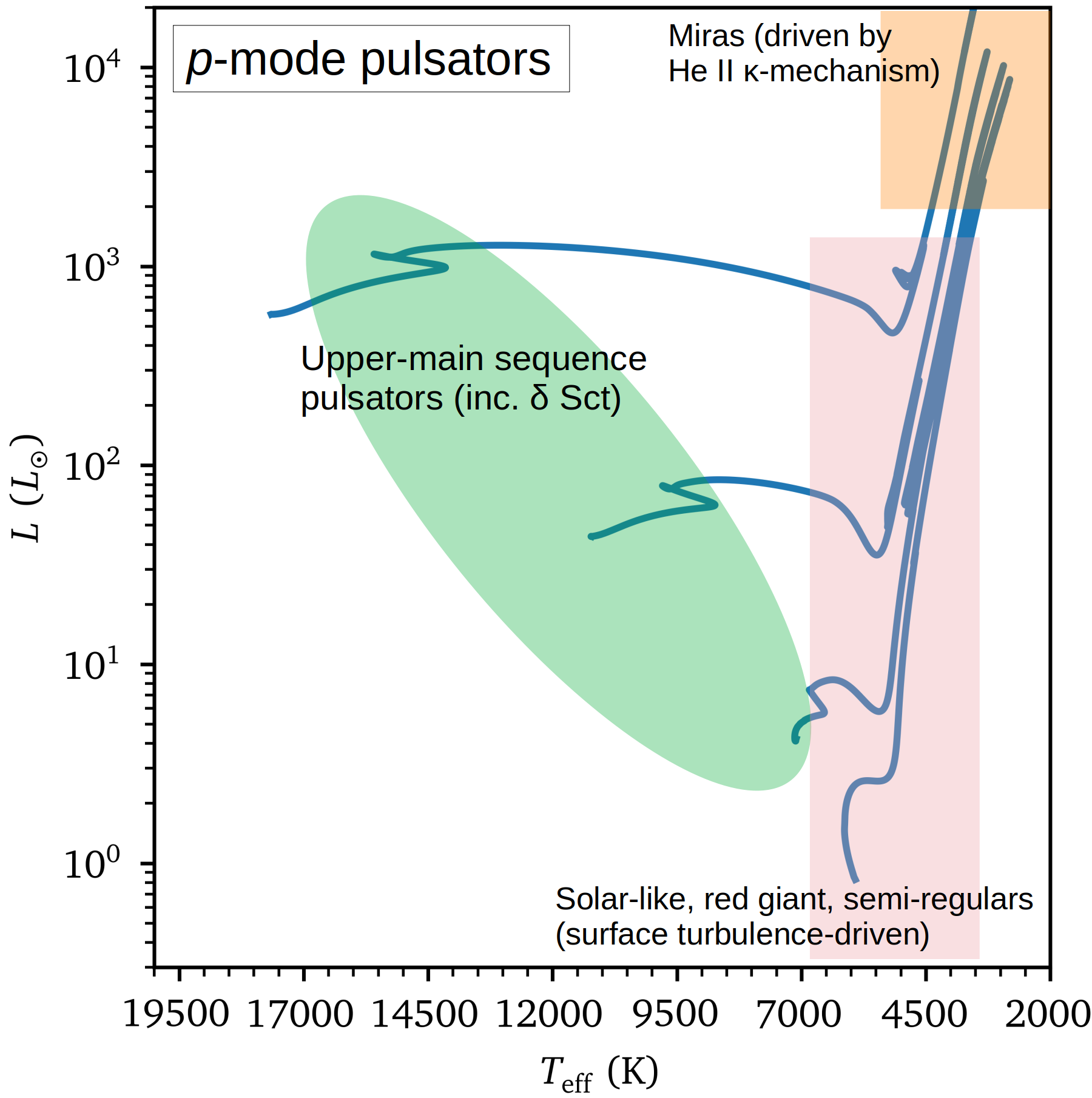}
\caption{H--R diagram showing the three types of p-mode pulsators investigated in this study. Tracks shown have solar metallicity and masses of 1.0, 1.4, 2.0, and 5.0$M_\odot$, from lower right to upper left.}
\label{fig:variability_HR}
\end{figure}
\input{tables/table1_inlist_summary}
Our analysis requires two self-consistent grids of seismic stellar evolutionary tracks computed under two different structural resolution assumptions, 
a means of defining evolutionarily-equivalent points along each pair of tracks, and 
definitions for the discrepancies in key asteroseismic quantities: frequency and period. We discuss each below.

\subsection{Stellar models}
We use \texttt{MESA} version 24.03.1 and \texttt{GYRE} version 7.2 (in adiabatic mode) to compute evolutionary and asteroseismic models. We focus on a small number of models representative of the types of stars most commonly studied with $p$-mode and mixed-mode ($p$ and $g$) observational asteroseismology. 
We include solar-mass models spanning four metallicities, $[\mathrm{Fe}/\mathrm{H}]=\{-2.0, -1.0, 0.0, +0.5 \}$, and three higher-mass models at solar composition, with masses of $1.4, 2.5$ and $5.0 M_\odot$. The upper main-sequence pulsators (including $\delta$ Scutis) are represented by main-sequence models with \Teff{} greater than 6800~K. The Miras are represented by models with $\log L$ greater than 3.15. Solar-like oscillators are represented by cool main-sequence models and low-luminosity red giants. Figure \ref{fig:variability_HR} shows the $p$-mode variability regimes covered by our grid. 

For all evolutionary calculations, we uniformly adopt a standard set of input physics, summarized in Table~\ref{table:inlist_summary}.\footnote{The complete inlists will be made publicly available at time of acceptance at [url TBD]}
The structural resolution in \texttt{MESA} refers to the number of radial (Eulerian) or mass (Lagrangian) coordinates, or grid points, used in the calculation of stellar profiles. This is also called the \texttt{mesh} resolution. By default, \texttt{MESA} uses adaptive mesh refinement (AMR), meaning the model is ``remeshed,'' or adjusted adaptively, from time step to time step \citep[][Section 6]{mesa2011}. While a user may require that a particular, fixed number of grid points be preserved throughout the evolution of the model, this is not the optimal technique for dealing with models that can change significantly in complexity during their evolution, or which require non-uniform sampling of the stellar interior.

Although the structural resolution can be controlled in this and a number of other ways, it is most commonly adjusted using the the \texttt{MESA} control \texttt{delta\_mesh\_coeff}.
This control does not itself dictate the mesh resolution; rather, it is a \textit{multiplier} on the default mesh resolution. By setting this control, the user specifies how much \textit{more} (or less) resolved they would like the model to be beyond the default resolution determined by \texttt{MESA}'s AMR routine.   
A value of \texttt{delta\_mesh\_coeff} = 1.0 corresponds to \texttt{MESA}'s default structural resolution, and a value of \texttt{delta\_mesh\_coeff} = 0.1 corresponds to $10\times$ the default structural resolution. As we will see, our results sharply demonstrate the inadequacy of using \texttt{MESA}'s default mesh resolution for most asteroseismic applications.

When the structural model is passed to \texttt{GYRE} to compute its frequency spectrum, we perform a standard re-meshing of the equilibrium model, adopting \texttt{GYRE} values of \texttt{w\_osc}= 10, \texttt{w\_exp}= 2 and \texttt{w\_ctr}= 10. The full set of \texttt{GYRE} namelist parameters is also provided with our \texttt{MESA} inlists. 
%  w_osc = 10 ! 10 Oscillatory region weight parameter
    % w_exp = 2  ! 2 Exponential region weight parameter
    % w_ctr = 10 ! 10 Central region weight parameter

\subsection{Equivalent evolutionary points}
To study the impact of structural resolution on frequency (period) predictions as a function of evolutionary phase requires defining a metric of evolutionary equivalence that can be maintained across tracks with different properties. We begin with the Equivalent Evolutionary Phase, or EEP, method outlined in \citet{Dotter2016}, which ensures consistent definitions of evolutionary phase across models with different physical or numerical properties.

A set of ``primary EEPs'' comprises the following critical points in stellar evolution, defined by an index $p$ and constructed such that the distance between two consecutive primary EEPs is 100:
\begin{itemize}
    \item[] \textbf{\boldmath EEP $p=0$, Protostar:} the initial point on the evolutionary track;
    \item[] \textbf{\boldmath EEP $p=100$, Zero-age main sequence:} the point at which the central hydrogen mass fraction, \texttt{center\_h1}, decreases by 0.0015 relative to its maximum value;
    \item[] \textbf{\boldmath EEP $p=200$, Intermediate-age main sequence:} the point at which \texttt{center\_h1} decreases to less than 0.3;
    \item[] \textbf{\boldmath EEP $p=300$, Terminal-age main sequence:} the point at which \texttt{center\_h1} decreases to less than $10^{-8}$;
    \item[] \textbf{\boldmath EEP $p=400$, Bottom of red-giant branch (RGB):} the point at which the mass coordinate at the base of the surface convection zone penetrates to within 75\% of the total stellar mass, or 0.25 $m/M_\odot$;
    \item[] \textbf{\boldmath EEP $p=500$, Onset of the RGB bump (if present):} the point at which the effective temperature (\Teff{}) begins to decrease along the RGB;
    \item[] \textbf{\boldmath EEP $p=600$, End of the RGB bump (if present):}  the point at which \Teff{} begins to increase again along the RGB, reversing direction after its descent;
    \item[] \textbf{\boldmath EEP $p=700$, Onset of helium flash:} the point at which the helium-burning luminosity $L_{\rm He}$ exceeds $10^{-4}$ of the total nuclear-burning luminosity;
    \item[] \textbf{\boldmath EEP $p=800$, Zero-age core-helium burning phase:} the point at which the central helium mass fraction \texttt{center\_he4} decreases by 0.0015 from its maximum value;
    \item[] \textbf{\boldmath EEP $p=900$, Terminal-age core-helium burning phase:} the point at which \texttt{center\_he4} decreases to less than 0.05
    \item[] \textbf{\boldmath EEP $p=1000$, Onset of the AGB bump (if present):} the point at which \Teff{} decreases for the first time along the asymptotic-giant branch (AGB);
    \item[] \textbf{\boldmath EEP $p=1100$, End of the AGB bump (if present):} the point at which \Teff{} begins to increase along the AGB after its first descent;
    \item[] \textbf{\boldmath EEP $p=1200$, Onset of the first thermal pulse on the AGB:} the point at which \texttt{center\_he4} falls below $10^{-20}$, and the hydrogen-to-helium burning luminosity ratio, $L_{\rm H} / L_{\rm He}$, exceeds 10;
    \item[] \textbf{\boldmath EEP $p=1200+(n-1)\times 100$: Onset of the $n$-th AGB thermal pulse:} subsequent episodes where $L_{\rm H} / L_{\rm He}$ temporarily drops below 10 and then rises above this threshold again.
\end{itemize}
This set of definitions differs somewhat from that presented in \citet{Dotter2016}, particularly in its denser coverage of the RGB and AGB phases. We do not consider the post-AGB or white dwarf cooling sequence here.

To enable the comparison of physical quantities at more precise locations, we next define sets of secondary EEPs between each pair of adjacent primary EEPs. 
We introduce an evolutionary distance metric, $D$, given by 
\begin{equation}
    D_{i+1} = D_{i} + \sqrt{\sum\nolimits_{j=1}^{N_j}w_j(x_{j,i+1}-x_{j,i})^2},
\end{equation}
where $D_i$ is the cumulative distance at the $i$-th point on the evolutionary track, $x_{j,i}$ is the $j$-th physical quantity at the $i$-th point, and $w_j$ are weighting factors applied to each physical quantity $x_j$, ensuring they are scaled appropriately for comparison. We adopt $w=\{2.5, 9.0\}$ with $x=\{\texttt{log\_center\_T}, \texttt{log\_center\_Rho}\}$ as default. 
The weightings come from forcing $\log \rho_\text{center}$ and $\log T_\text{center}$ to a normalized scale, where $\log \rho_\text{center}$ ranges from $-3$ to 6 across our models and $\log T_\text{center}$ ranges from 6.0 to 8.5.
To ensure that our results are not sensitive to the choice of metric, we tested an alternative definition of $D$, adopting $w=\{0.9, 4.6\}$ with $x=\{\texttt{log\_Teff}, \texttt{log\_L} \}$. This yields no significant differences in our results.

To ensure that our error estimates are not dominated by time resolution, we interpolate all physical quantities as functions of stellar age for each evolutionary track and resample the models at a tenfold higher age resolution. We also tested a hundredfold resampling, which did not produce significant differences compared to the tenfold resampling.

We compute $D_i$ for every resampled model (time step). The secondary EEPs $\{s_0...s_{100}\}$ are placed by 
min-max scaling 
$D$ from 0 to 100 across each region demarcated by two adjacent primary EEPs:
\begin{equation}
     s =  [D - \mathrm{min}(D) / (\mathrm{max}(D) - \mathrm{min}(D))] \times 100.
\end{equation}
The final EEPs we use as reference points for comparison between tracks with different resolutions are the result of adding the primary EEP $p$ to the secondary EEP $s$.
For example, if we are inspecting frequencies at secondary EEP $s=53$ between the terminal-age main sequence ($p=300$) and the bottom of the red giant branch ($p=400$), the reference EEP is $p+s = 353$.

In subsequent analysis, we measure the frequency discrepancy at six representative evolutionary points on each track: 
the main sequence (MS) at EEP 187, 
the subgiant branch at EEP 303, 
the early red giant branch (RGB) at EEP 410,
the late RGB at either EEP 665 (if the RGB bump is present) or 495 (if the RGB bump is not present), 
the core helium-burning phase (CHeB) at EEP 850, and
the asymptotic giant branch (AGB) at EEP 1198.
These henceforth serve as reference points for detailed examination of the models.

\begin{figure*}
    \centering
    \includegraphics[width=1.0\linewidth]{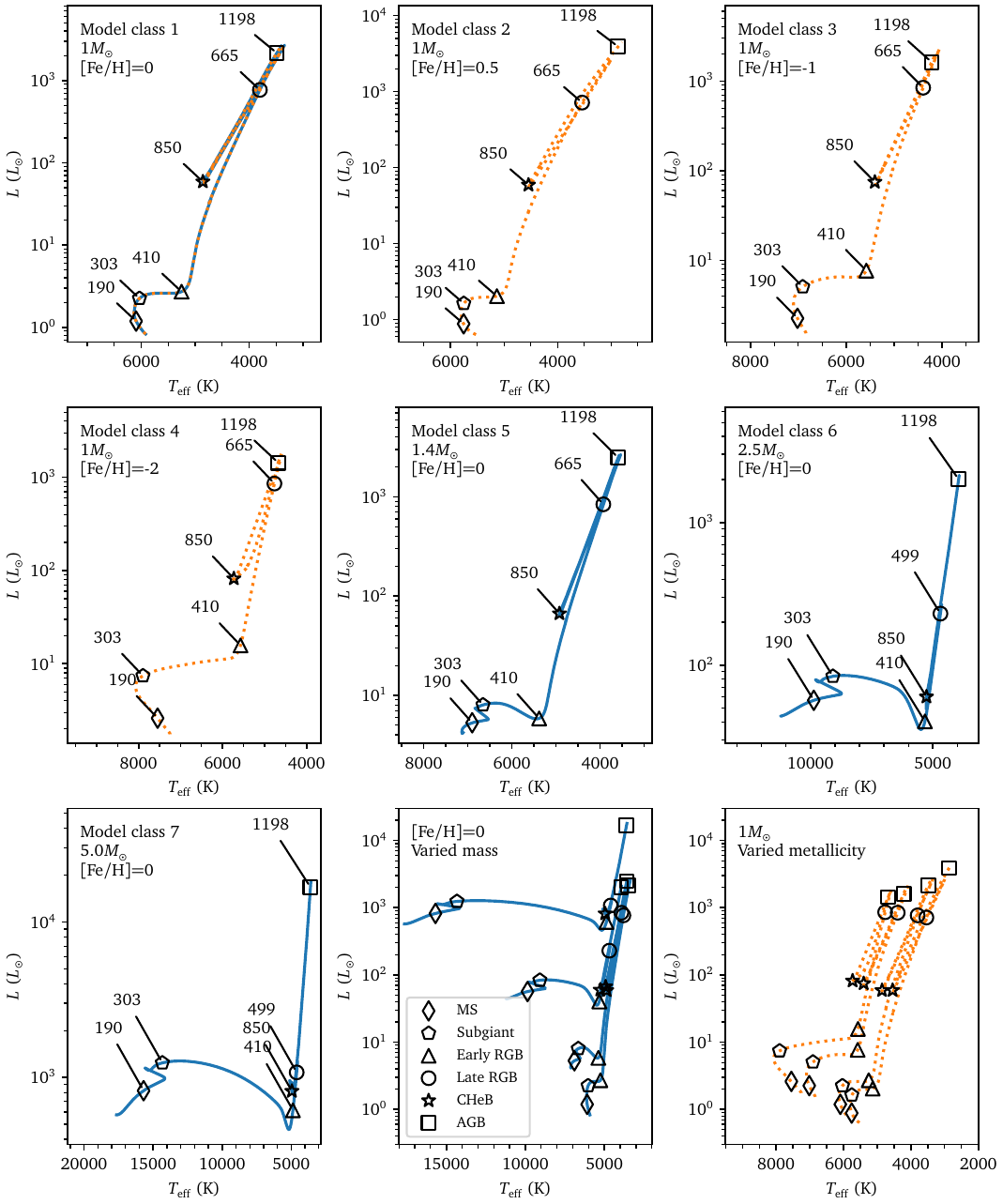}
    \caption{Panels 1 through 7 show the locations and indices of Equal Evolutionary Points (EEPs) for each model class, with the class indicated in the upper left. The ages and other fundamental parameters measured at these EEPs are summarized in Table \ref{table:overview}. }
    \label{fig:eep_HR}
\end{figure*}

We define seven classes of models, each corresponding to a particular physical configuration adopted in the evolutionary track. Model Class 1 contains a main sequence, subgiant, early RGB, late RGB, core helium-burning, and AGB model with solar mass and solar metallicity. Model Class 2 assumes solar mass and a metal-enriched composition of $[\mathrm{Fe}/\mathrm{H}]=+0.5$\,dex, representing the approximate upper limit on metal-enrichment in the Milky Way. Model Classes 3 and 4 use solar mass and metallicities of $[\mathrm{Fe}/\mathrm{H}]=-1.0$, and $[\mathrm{Fe}/\mathrm{H}]=-2.0$, respectively. 
Model Classes 5, 6, and 7 adopt solar composition and masses of $1.4$, $2.5$, and $5.0 M_\odot$, respectively. 
Table~\ref{table:overview} provides a summary of critical stellar parameters---age, T$_\text{eff}$, $\log L$, and $\log g$---at each reference EEP and corresponding evolutionary phase within each model class. 
Figure~\ref{fig:eep_HR} shows the reference EEPs
for each model class on the H--R diagram.

\subsection{Differences in Frequency and Period}

\input{tables/table2_overview_model_classes}

To describe the \textit{absolute frequency discrepancy} between two points on models computed with identical input physics but different structural resolutions,
we introduce the quantity $\dm$:
\begin{multline}
\label{eqn:dm}
     \dm (\text{EEP}) 
	\\ = | \nu_{\texttt{[dmesh=1.0]}}(\text{EEP}) - \nu_{\texttt{[dmesh=0.1]}}(\text{EEP}) |,
\end{multline}
where $\nu_{\texttt{[dmesh=1.0]}} (\text{EEP})$ refers to the test frequency $\nu$ measured at EEP on a track adopting the default (low) mesh resolution, \texttt{delta\_mesh\_coeff} = 1.0, and $\nu_{\texttt{[dmesh=0.1]}} (\text{EEP})$ refers to the frequency measured at EEP on the otherwise physically identical evolutionary track adopting $10\times$ the default structural resolution, \texttt{delta\_mesh\_coeff} = 0.1.

To determine which test frequency (mode) should be used to calculate $\dm$, we select frequencies in each domain representative of the modes most likely to be observed in stars of that type. We use $p$-mode frequencies for modes adjacent to \numax{} for solar-like oscillators, 
the $n=5$ mode for upper main-sequence pulsators, and 
the fundamental ($n=1$) mode for Miras. For all mode choices, spherical degree and azimuthal order are $l=0$ and $m=0$, respectively. 
The values of \numax{} are calculated as \citep{Brown1991,KB95}
\begin{multline}
\label{unused}
    \numax
	\\ = 
    \left(\frac{M}{M_\odot}\right) 
    \left(\frac{R}{R_\odot} \right)^{-2} 
    \left(\frac{\Teff}{5777\,{\rm K}}\right)^{-0.5} 
    \times {3090\,\mu{\rm Hz}}.
\end{multline}

The \dm{} quantity is calculated at the preferred EEPs summarized in Table \ref{table:overview}.
For upper main-sequence and Mira oscillators, we use cubic interpolation to estimate the mode frequency for $n = 3$ or $n = 1$ as a function of EEP. 
For solar-like oscillators, the process involves three steps. First, cubic interpolation is used to estimate the mode frequencies of four radial ($n$) orders that bracket \numax{}, as a function of EEP. Second, \dm{} is evaluated for each of these orders at the desired EEPs (187, 303, 410, etc). Finally, cubic interpolation is performed again on \dm{} as a function of mode frequency $\nu$, and the final reported \dm{} is evaluated at \numax{}.

For longer-period variables, such as more massive red giants and AGB stars, it is more useful to express the discrepancy in terms of period rather than frequency.
We thus introduce an analogous quantity representing the \textit{absolute period discrepancy}, $\dPm$, defined as
\begin{multline}
\label{eqn:dPm}
    \dPm (\text{EEP}) 
	\\
	= | P_{\texttt{[dmesh=1.0]}} (\text{EEP}) - P_{\texttt{[dmesh=0.1]}} (\text{EEP}) |.
\end{multline}
The computation procedure is analogous to that of \dm{}, except periods are used in the interpolations instead of frequencies.

\section{Resolution Uncertainty Calculations}
\label{sec:results}

\input{tables/table3_solar-like}

\input{tables/table4_dSct-like}

\input{tables/table5_Miras}
\begin{figure*} 
\includegraphics[width=2\columnwidth]{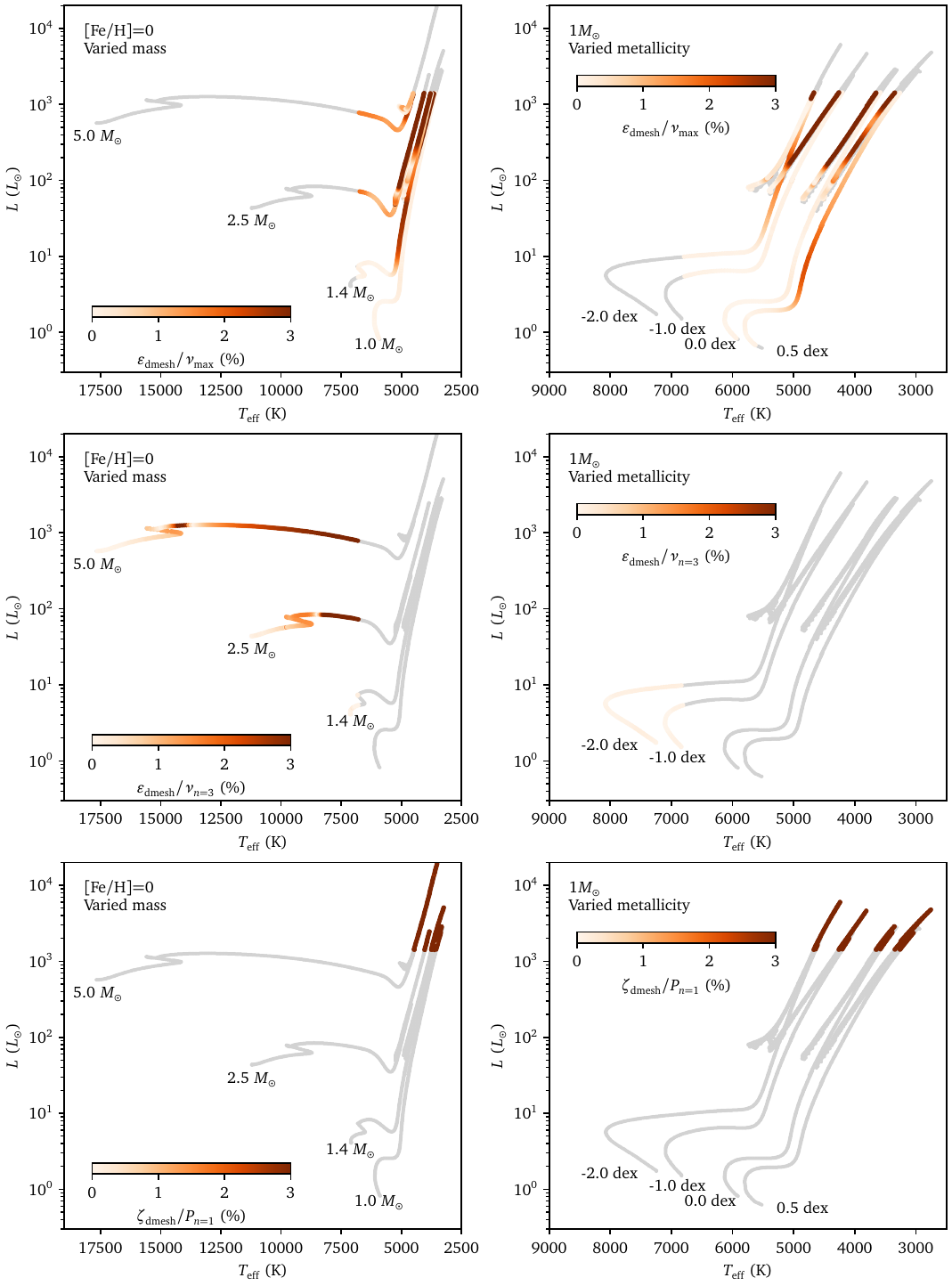}
\caption{HR diagram color-coded by $\dm$ (\dPm{}), the degree of discrepancy in representative test frequency $\nu_\text{test}$ ($P_\text{test}$). The left panel shows all models with solar metallicity but different masses, and the right panel with solar mass but different metallicites. 
}
\label{fig:dmesh}
\end{figure*}
We present calculations of the resolution uncertainty term for solar-like, upper main-sequence, and Mira oscillators in
Tables \ref{table:solar-like}, \ref{table:dSct}, and \ref{table:Miras}, respectively, for the set of reference EEPs representing different evolutionary phases introduced in Section \ref{sec:methods}. 
%(see Table \ref{table:overview}).  
%
Tables \ref{table:solar-like} and \ref{table:dSct} provide the frequency-based resolution uncertainty term $\dm$ (equation \ref{eqn:dm}), whereas Table \ref{table:Miras} presents the analogous quantity, $\dPm$, in period (equation \ref{eqn:dPm}).

Table \ref{table:solar-like} shows $\dm$ calculations for solar-like oscillators, whose pressure modes are driven by stochastic excitation in the surface convection zone. The most useful and likely-to-be-observed $p$-mode in this case is the one closest to $\nu_\text{max}$, the frequency of maximum power. As $\nu_\text{max}$ changes depending on the properties of the solar-like oscillator, so does our test frequency. Table \ref{table:solar-like} thus provides the radial orders $n$ of the two modes bracketing $\nu_\text{max}$ along with $\nu_\text{max}$, $\dm$, and the quantity $\dm/\nu_\text{max}$, in $\mu$Hz. The lattermost of these quantities indicates the fractional uncertainty.

Table \ref{table:dSct} shows the same for upper main-sequence oscillators, but the test frequency in these cases is the one corresponding to a fixed mode of radial order $n=3$ ($\ell, m = 0$).
This was chosen because $\delta$ Scutis typically pulsate in low-order p modes, although the precise mode selection process remains somewhat uncertain \citep{Dupret2005,Xiong2016}.
Frequencies are reported here in cycles per day.

Lastly, Table \ref{table:Miras} presents the resolution-induced discrepancy in terms of period, in days, for Miras, using a test period $P_\text{test}$ fixed to $n=1$ ($\ell,m=0$) and the absolute period discrepancy $\dPm$. Pressure variations in this regime are driven by the $\kappa$-mechanism in the He II ionization zone.
The most appropriate test value here is the period corresponding to the fundamental pressure mode, or FM, as this is the mode overwhelmingly likely to correspond to the dominant periodicity observed in these stars
\citep{Soszynski2013, Trabucchi2017,Molnar2019,Trabucchi2021,Joyce2024}.

Figure \ref{fig:dmesh} shows the evolution of the fractional uncertainty term across the HRD. Panels in the left-hand column show tracks with the same metallicity and different masses; panels in the right-hand column show tracks with the same mass and different metallicities. Panels in the top row highlight solar-like oscillators, panels in the middle row show upper main-sequence oscillators, and panels in the bottom row show Miras.

Notable features and trends revealed by Tables \ref{table:solar-like}, \ref{table:dSct} and \ref{table:Miras} and Figure \ref{fig:dmesh} include: 
\begin{itemize}
    \item[-] resolution-based uncertainty contributions $\dm/\nu$ and $\dPm/P$ are larger for higher-mass models and higher-metallicity models;
    \item[-] $\dm/\nu$ and $\dPm/P$ generally increase as evolutionary stage increases, with the AGB/Mira regions hosting contributions near the 3\% level;
    \item[-] fractional uncertainties for solar-like oscillators are below 1\% for models $\le 1.4M_\odot$; and
    \item[-] AGB models show the largest variance in $\dPm/P$---between 0.5 to 20\%---across mass and metallicity and the largest absolute error contribution, with $\dPm/P_\text{test}$ near or exceeding 10\% in four of the seven model classes. 
\end{itemize}
While it is generally well-established that modeling uncertainties of all types become larger the further the model deviates from solar parameters 
\citep{Joyce2018-mp,Tayar2022,Cinquegrana2022,Joyce2023,Cinquegrana2023}, these calculations provide quantitative benchmarks as a function of evolutionary phase for the (un-)reliability of stellar models computing with \texttt{MESA}'s default resolution setting.

\section{Comparison to \textit{TESS} Observational Uncertainties}
\label{sec:TESS}

\input{tables/table6_compare_to_TESS_solar}

\input{tables/table7_compare_to_TESS_dSct}

\input{tables/table8_compare_to_TESS_Mira}
\begin{figure*} 
\includegraphics[width=2\columnwidth]{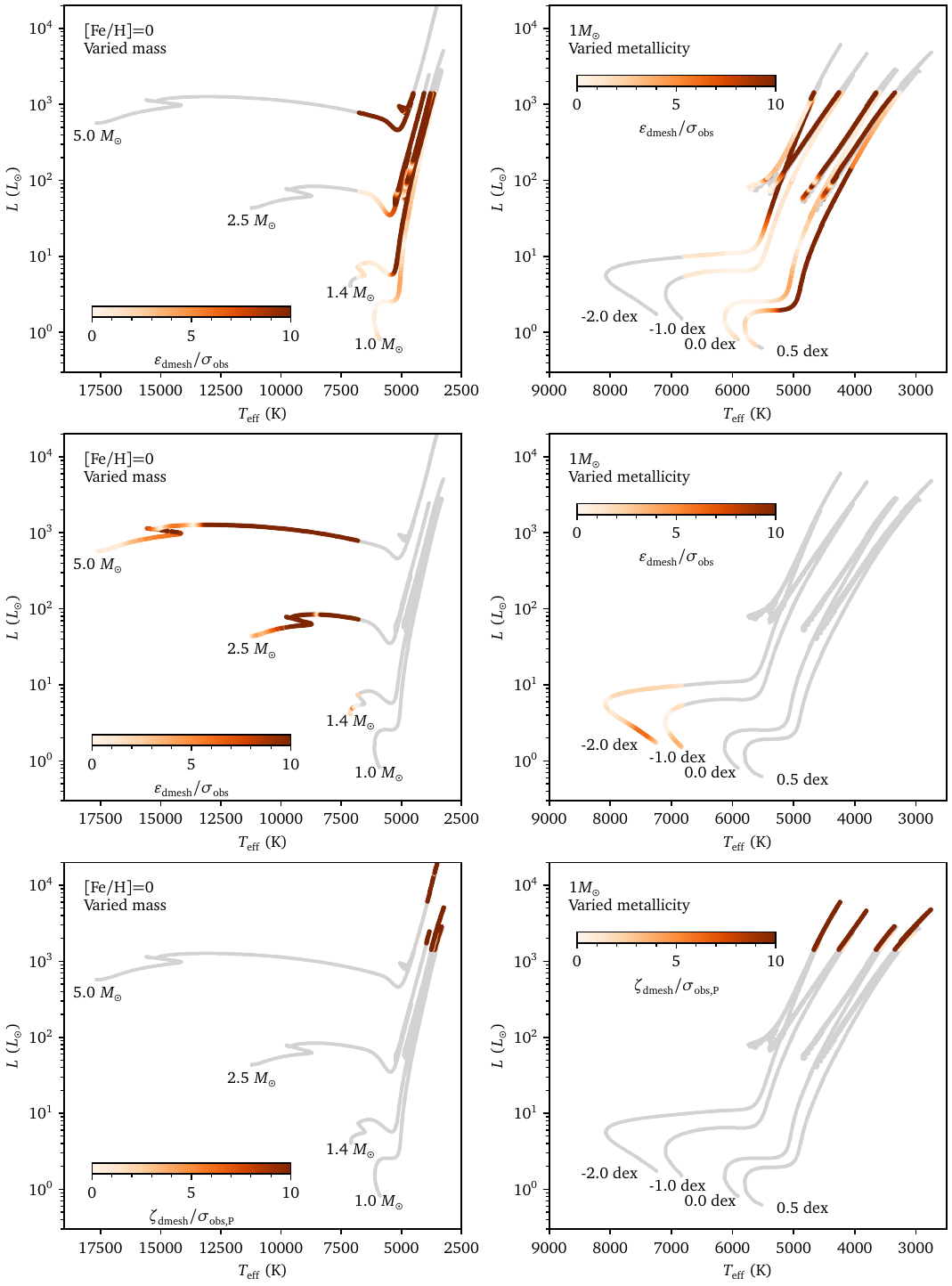}
\caption{H--R diagram color-coded by the ratio of \dm{} (\dPm{}) to the typical observational uncertainties $\sigma_\text{obs}$ (or $\sigma_\text{obs, P}$), assuming the worst-case observing scenario (T=1 month, $\sigma_\text{noise}=197$ ppm/hr). The left panel shows all models with solar metallicity but different masses, and the right panel with solar mass but different metallicites.}
\label{fig:TESS_ratio}
\end{figure*}
With \% $\dm$ and \% $\dPm$ falling below the 1\% level for the majority of models, it is reasonable to ask how important such considerations are in the context of other sources of uncertainty in the asteroseismic parameter determination process. To answer this question, it is most informative to compare the mesh resolution uncertainties calculated above to typical observational uncertainties as a function of stellar type. This section  uses \TESS{} \citep{Ricker2015} as the example instrument, though these calculations can be easily translated to \Kepler{} \citep{Borucki2010}, and the method can be generalized to any instrument.

\subsection{Simulating Observational Uncertainties}
We now simulate the typical uncertainty for observed oscillation modes of the types studied here (solar-like, $\delta$ Scuti, and Mira) using the noise characteristics of the \TESS{} instrument.

To represent realistic observational scenarios, we select two noise values corresponding to 5th and 10th magnitude stars. Assuming a nominal 2~min cadence, the per-cadence noise $\sigma_{\rm noise/cadence}$ is determined using the ATL calculator provided by \citet{Hey2024}\footnote{\url{https://github.com/danhey/tess-atl}}.  
Regardless of the chosen cadence, the time-domain scatter $\sigma_{\rm noise}$ is approximately 19~ppm$\cdot$hr$^{-1}$ and 197~ppm$\cdot$hr$^{-1}$, respectively.
These values, without loss of generality, are comparable to those for 10th and 15th magnitude stars observed by \Kepler{} \citep{Gilliland2011}.

We simulate mode amplitudes based on stellar parameters. For solar-like oscillators, we adopt established scaling relations for predicting asteroseismic detection yields:
\begin{equation}
    A_{\rm osc} = \frac{(L/\Lsun)^{0.84} \beta }{(M/\Msun)^{1.32} (\Teff/{\rm K}) c_k} \times 22167.48 ~{\rm ppm},
\end{equation}
where $c_k = (\Teff/5934 \ {\rm K})^{0.8}$ \citep{Huber2011,Chaplin2011,Campante2016,Schofield2019}.
The factor $\beta$ accounts for amplitude suppression for stars near the $\delta$ Sct instability strip,
\begin{equation}
    \beta = 1 - \exp((\Teff - T_{\rm red} (L/\Lsun)^{0.093})/1550\ {\rm K})
\end{equation}
\citep{Schofield2019,Hey2024}.
For upper main-sequence pulsators, we adopt a conservative value of $\Aosc=14$~ppm, reflecting the typical range of $\delta$ Sct oscillation amplitudes, which span from 14 to 30000 ppm \citep{aerts}. 

For Mira variables, we applied an amplitude--luminosity relation calculated based on the high-luminosity variable catalog of  \citet{Yu2020}:
\begin{multline}
\label{eq:amp-lum}
    \log (A_{\rm osc}/{\rm ppm}) 
    \\
    = -0.82 \log (L/\Lsun) + 0.45 \log(L/\Lsun)^2 + 2.00.
\end{multline}

Next, we calculate mode lifetimes. Solar-like oscillations have finite lifetimes $\tau$, which are related to mode linewidths $\Gamma$ through $\tau = 1 / (\pi \Gamma)$. Mode linewidths are determined using the fit equation:
\begin{multline}
\label{unused3}
    \log(\Gamma/\Gamma_0)
    \\
    = c_1\log(\Teff/5777\ {\rm K}) + c_2\log(g/274~{\rm m\cdot s^{-1}}),
\end{multline}
where $\Gamma_0$, $c_1$, and $c_2$ are coefficients provided in Table 5 of \citet{Li2020}. We adopt separate parameterizations for main-sequence (MS) and red giant branch (RGB) stars, with a \logg{} cutoff of 3.5 distinguishing the two regimes. 
For $\delta$ Sct and Mira-like variables, the modes were treated as coherent with effectively infinite lifetimes.

Based on these noise properties, mode amplitudes, and mode lifetimes, we estimate the expected uncertainties of mode frequencies as in \citet{kb12}:
\begin{equation}
    \sigma_{\rm obs} \approx 0.44 \sqrt{\pi/N}\sigma_{\rm noise/cadence}A_{\rm osc}^{-1} \sqrt{T^{-2} + \tau^{-2}},
\end{equation}
where $T$ is the total observing time and $N$ is the number of data points in the time series (see also \citealt{Libbrecht1992,Ballot2008,Lund2017}). 
In period, this becomes
\begin{equation}
    \sigma_{{\rm obs}, {P}} = \frac{\sigma_{\rm obs}}{\nu_{\rm test}^2}.
\end{equation}
We consider three typical values for the total observing time $T$: 4 years, 1 year, and one month (0.0833 years).

\subsection{Ratio Between Numerical And Observational Uncertainties}

Tables~\ref{table:TESS_solar} (solar-like), \ref{table:TESS_dSct} (upper main-sequence), and \ref{table:TESS_Mira} (Miras) compare resolution-based numerical uncertainty to observational uncertainty
for six different combinations of observing strategies, defined by three observing durations ($T$ = 1 month, 1 year, and 4 years) and two noise levels ($\sigma_{\rm noise}$ = 19.1 ppm/hr and 197 ppm/hr), which corresponds to 5th and 10th magnitude TESS stars or 10th and 15th magnitude \Kepler{} stars.
The ratio $\dm$/$\sigma_\text{obs}$ (or $\dPm$/$\sigma_\text{obs, P}$) provides a measure of the numerical uncertainty induced by mesh resolution relative to observational noise.
As in Tables \ref{table:solar-like}, \ref{table:dSct} and \ref{table:Miras}, measurements are taken at preferred EEPs representative of different stellar evolutionary phases. 

When assuming the best-case observing scenario, with a 4-year observing time and a noise level of 19.1 ppm/hr, we find that resolution-based numerical uncertainties are consistently orders of magnitude larger than the observational uncertainties, regardless of evolutionary phase or stellar type. The situation is least dire for solar-like oscillators, whose resolution-based uncertainties dominate over observational uncertainties by a factor of a few hundred; upper main-sequence models are 10 to 100x worse than this, and Miras are another factor of ten worse than $\delta$ Scutis.

Following Figure \ref{fig:dmesh}, we compute the observational uncertainty $\sigma_{\rm obs}$ under the assumptions of the worst-case observing scenario ($T=0.038$ yr, $\sigma_\text{obs} = 197$) and evaluate the ratio of $\dm$ to $\sigma_{\rm obs}$ (or $\dPm$ to $\sigma_\text{obs, P}$) along the entire evolutionary tracks representing each model class.
The worst-case observing scenario produces mesh-to-observational uncertainty ratios of order unity for most solar-like oscillators, indicating that $\dm$ is comparable to the observational uncertainty in this scenario. As expected per the calculations in Section 3, these ratios are worse for higher-mass stars and later evolutionary stages. Ratios for $\delta$ Scutis are comparable to those for solar-like oscillators under worst-case observing assumptions, and ratios for Miras are about a factor of 10 worse.

It is worth noting that in the case of Mira variables, the observational uncertainty as we have computed here---based primarily on instrumentation---is very unlikely to be among the dominant sources of uncertainty in characterizing them observationally. Miras are notoriously difficult to constrain observationally due issues such as meandering, period decay, and large-amplitude brightness variations, all of which can shift the values of their fundamental parameters significantly over relatively short periods of time \citep{Kiss2006, Banyai2013, Molnar2019, Merchan-Benitez2023}. As many studies have shown, modeling such stars is similarly complicated \citep[e.g.][]{Karakas2011}, and for that reason, proper AGB modeling demands very high structural resolution and treatment of non-adiabatic effects \citep[e.g.][]{Zinn2023, Joyce2024}. While we do not incorporate non-adiabatic effects, mass loss, or other considerations necessary for proper AGB modeling here, the extreme values of $\dPm$/$\sigma_\text{obs, P}$ in Table \ref{table:TESS_Mira} nonetheless provide a representative portrait of the extent to which modeling uncertainty obliterates instrumental noise for late-stage, high-amplitude variable stars.

\section{Implications for Modern Asteroseismic Techniques}
\label{sec:impacts}
One of the key powers of asteroseismology is its ability to provide constraints on the sub-surface properties of stars. This enables far more precise determinations of the fundamental stellar parameters that are closely tied to the physics of stellar interiors, such as mass and age,
than can be obtained using classical observational techniques alone.

So far, we have quantified the impact of mesh uncertainty on individual $p$-mode frequencies. While valuable in their own right, these frequencies are, more importantly, essential ingredients in a number of modern asteroseismic techniques 
ranging from period-spacing diagrams to age determination methods. We now discuss the potential impacts of improper treatment of structural resolution in the context of these techniques. We begin with the most direct inferences based on $p$-modes and conclude with the most derived.

%%%%%%%%%%%%%%%%%%%%%%%%%%%%%%%%%%%%%%%%%%%%%%%%%%
% 5.1
%%%%%%%%%%%%%%%%%%%%%%%%%%%%%%%%%%%%%%%%%%%%%%%%%%
\subsection{Evolutionary Features on the H--R Diagram}

\begin{figure}
    \centering
    \includegraphics[width=1.0\linewidth]{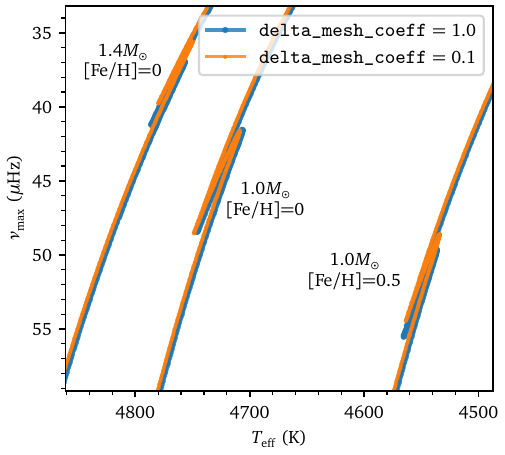}
    \includegraphics[width=1.0\linewidth]{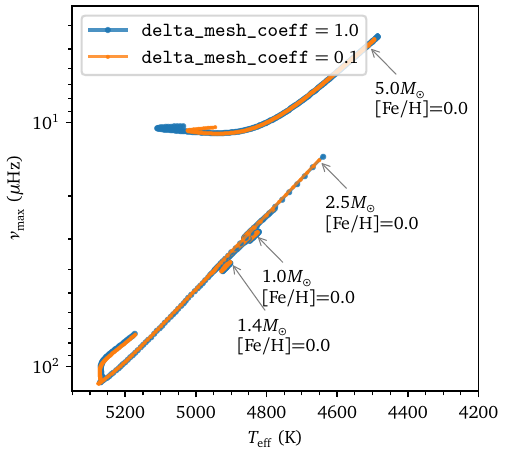}
    \caption{Seimic H--R diagram (\numax{} vs. \Teff{}) around the RGB bump (top panel) and the CHeB phase (bottom panel). }
    \label{fig:features}
\end{figure}

The red giant branch bump, or RGBB, is an evolutionary feature that occurs in stars with surface convective envelopes. During this phase, the star's luminosity temporarily drops along the RGB as a result of the encounter between the outwardly expanding hydrogen-burning shell and the chemical discontinuity left by the receding convective envelope \citep{Joyce2015,JCD2015,Tayar2022-rgbb}. 
Because the RGBB is a direct probe of the maximum depth of penetration of the convective envelope \citep[e.g.][]{Fraser2022}, convective overshoot in the envelope reduces the luminosity, or increases the oscillation frequency, \numax{}, at which the bump occurs. This makes $p$-modes a sensitive probe of the physics of envelope overshoot \citep{Angelou2015,Khan2018}. 

The upper panel of Figure~\ref{fig:features} illustrates the impact of \dm{} on the location of the RGBB in the \numax{}-\Teff{} plane for tracks with convective envelopes.
The median fractional uncertainty of \numax{} is 0.6\%
\citep{Pinsonneault2024}, yet we observe differences of approximately 2\% in the locations of the maximal and minimal luminosities of the bump due to differences in \dm{}. The resolution-induced uncertainty therefore dominates over the fractional uncertainty of \numax{} by more than a factor of three.

The locations of core helium burning (CHeB) and the horizontal branch (red clump) more broadly are influenced by core overshoot, both during the main sequence (i.e.\ relic effects) and during the clump phase, and are therefore also sensitive to mesh resolution.
Figure~\ref{fig:features} demonstrates the influence of \dmesh{} on evolutionary tracks of varying mass in the \numax{}–\Teff{} diagram. The agreement for high-mass tracks is poorer, with temperature differences reaching up to 100~K. Additionally, morphological variations are evident, stemming from different resolution treatments, especially near the convective core boundary.

Under-resolving the stellar profile will always impact a model's evolution to some degree, but it is especially dangerous around boundary regions. For example, the Schwarzschild or Ledoux criterion for stability against convection is evaluated at every mesh point in a stellar model, and this determines the location, size, and persistence of convective and radiative zones. 

The development of convection zones in models is sensitive to resolution, and whether a model star develops a convective core on the main sequence significantly impacts its subsequent evolution. Similarly, any model that hosts a complicated interior structure, such as the double-shell burning in AGB stars, will be particularly sensitive to mesh resolution. Modeling AGB stars requires adequately resolving multiple boundaries, and the number and duration of thermal pulses an AGB model undergoes is well known to be sensitive to the number of grid points used in the interior \citep{Cinquegrana2023,Joyce2024}.

%%%%%%%%%%%%%%%%%%%%%%%%%%%%%%%%%%%%%%%%%%%%%%%%%%
% 5.2
%%%%%%%%%%%%%%%%%%%%%%%%%%%%%%%%%%%%%%%%%%%%%%%%%%
\subsection{Seismic Parameters: Period Spacings, Frequency Ratios, and Phase Offset }

\begin{figure}
    \centering
    \includegraphics[width=1.0\linewidth]{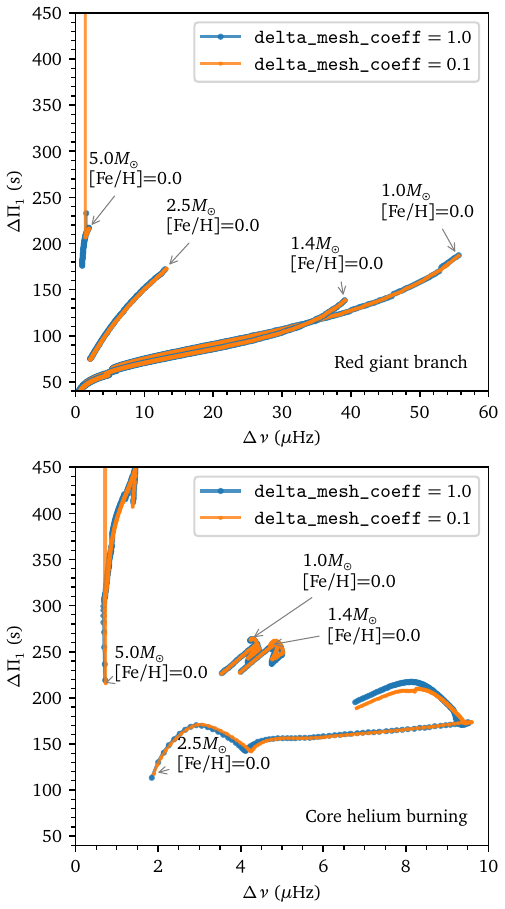}
    \caption{Seismic \DPi{1}--\Dnu{} diagram for RGB (top panel) CHeB phase (bottom panel).}
    \label{fig:Dnu_DPi}
\end{figure}

%%%%%%%%%%%%%%%%%%%%%%%%%%%%%%%%%%%%%%%%%%%%%%%%%%
% 5.2.1
%%%%%%%%%%%%%%%%%%%%%%%%%%%%%%%%%%%%%%%%%%%%%%%%%%

In Section~\ref{sec:results}, we analyzed the frequency response to \dmesh{} as a function of evolutionary phase (EEP). However, evolutionary stage is not directly measurable through observation. Instead, seismic observables, such as the $p$-mode frequency separation (\Dnu{}), $g$-mode period spacing (\DPi{1}), frequency ratios ($r_{02}$ and $r_{13}$), and the $p$-mode phase offset ($\epsilon_p$) are more commonly compared to predictions from stellar models. 
We should therefore assess the impact of \dmesh{} in terms of these seismic observables. In this analysis, we focus on the variation of seismic parameters as a function of the large frequency separation, \Dnu{}, calculated from linearly fitting radial mode frequencies $\nu_{n,\ell=0}$ as a function of radial order $n$:
\begin{equation}\label{eq:asymp}
    \nu_{n,\ell=0} = \Dnu{} ( n + \epsilon_p ).
\end{equation}

The $g$-mode period spacing is given by
\begin{equation}
    \DPi{l} = \frac{2\pi^2}{\sqrt{l(l+1)}}\left( \int \frac{N}{r} {\rm d} r \right)^{-1},
\end{equation}
where $N$ is the buoyancy frequency. This quantity probes the inner cavity of the near-core, radiative region in which $g$ modes propagate 
\citep{Pedersen2021,Noll2024}. 
Figure~\ref{fig:Dnu_DPi} compares the evolution of \DPi{1} as a function of \Dnu{} for tracks with different resolutions. The upper panel highlights the red giant region, and the lower panel highlights the core helium-burning region. 
In both regions, higher-mass models show the largest discrepancy (i.e.\ poorest convergence). This is because \DPi{1} is measured from the convective core boundary, which will be improperly resolved with low mesh sampling.
The typical uncertainty in observational measurements of \DPi{1} is below 1~s \citep{Vrard2016,Gehan2018,Kuszlewicz2023,Lig2024,Hatt2024}. While the change in \DPi{1} as a function of \texttt{delta\_mesh\_coeff} is below this threshold for the 1.0 and 1.4$M_{\odot}$ tracks in both the red giant and core helium-burning regions, the resolution-based discrepancy for the higher-mass (2.5, 5.0$M_{\odot}$) tracks ranges from 2-10 seconds.

%%%%%%%%%%%%%%%%%%%%%%%%%%%%%%%%%%%%%%%%%%%%%%%%%%
% 5.2.2
%%%%%%%%%%%%%%%%%%%%%%%%%%%%%%%%%%%%%%%%%%%%%%%%%%
% \subsection{\texorpdfstring{\Dnu{}}{Dnu}--\texorpdfstring{\dnu{}}{dnu} diagram}
\begin{figure}
    \centering
    \includegraphics[width=1.0\linewidth]{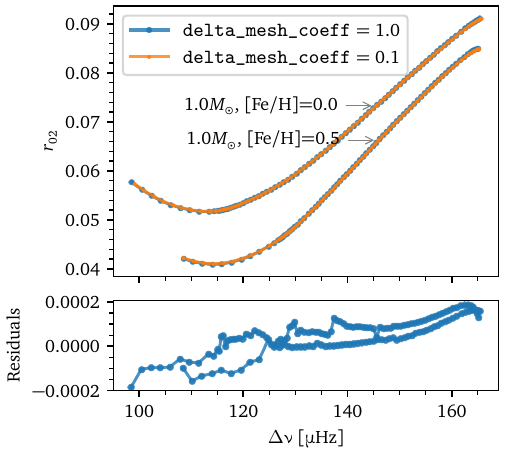}
    \includegraphics[width=1.0\linewidth]{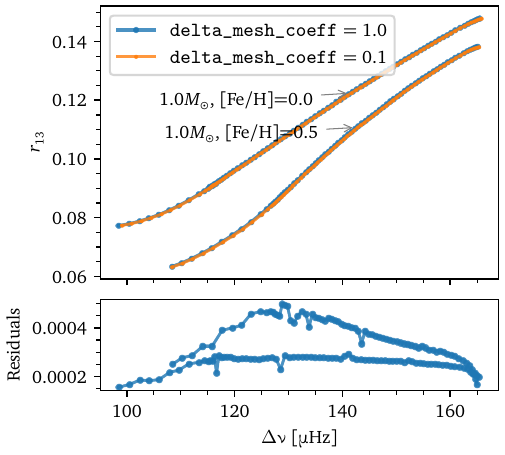}
    \caption{Seismic ratios ($r_{01}$ and $r_{13}$) vs. \Dnu{} for low-mass main-sequence models. The bottom panels show the residuals between the two \dmesh{} tracks.}
    \label{fig:Dnu_r02_r13}
\end{figure}

\begin{figure}
    \centering
    \includegraphics[width=1.0\linewidth]{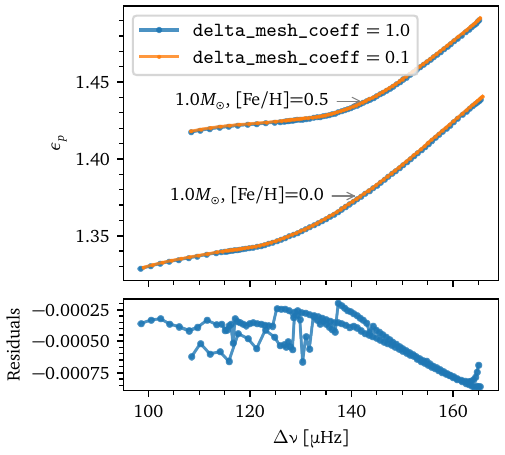}
    \caption{P-mode phase offset $\epsilon_p$ vs. \Dnu{} for low-mass main-sequence models. The bottom panels show the residuals between the two \dmesh{} tracks.}
    \label{fig:Dnu_eps}
\end{figure}

For low-mass and solar-like main-sequence stars, seismic frequency spacing ratios such as $r_{02}$ and $r_{13}$ can be used to estimate stellar ages \citep[e.g.][]{Joyce2018-acen} because they are sensitive to the near-core composition \citep{JCD1984,Roxburgh2003}. 
Likewise, the $p$-mode phase term, $\epsilon_p$, aids in mode identification (e.g., in F stars and $\delta$ Scuti stars; \citealt{White2011,Murphy2021}) and can probe the acoustic depths of the HeII ionization zone and the bottom of the outer convection zone \citep{Broomhall2014,Verma2019a,Dreau2020}.
We calculated frequency ratios following the definitions of \citet{Roxburgh2003} and averaged across radial orders, and calculated $\epsilon_p$ following Equation~\ref{eq:asymp}.
Figures~\ref{fig:Dnu_r02_r13} and~\ref{fig:Dnu_eps} present frequency ratios and $\epsilon_p$, respectively, as a function of \Dnu{} for solar-mass tracks. Per the residuals in the lower panels of these Figures, differences are minimal when these three quantities are evaluated at the same \Dnu{} but different \texttt{delta\_mesh\_coeff}. 
Hence, these quantities are robust against resolution uncertainty.

It is important to recognize that \dmesh{} influences both \Dnu{} and the seismic quantities discussed above in a correlated manner. Hence, structural changes in the star induced by \dmesh{} will propagate through all mode frequencies and therefore all of these quantities, leading to interdependent variations. As a result, the projections of these variations along seismic axes (\Dnu{})
may make discrepancies appear smaller than they actually are as a function of evolutionary stage (i.e.\ in EEP space). 

Though we do not investigate this issue explicitly here, it is further worth noting that mixed modes, being the result of coupling between $p$ and $g$ modes, will be affected by $p$ mode uncertainties as well.

% %%%%%%%%%%%%%%%%%%%%%%%%%%%%%%%%%%%%%%%%%%%%%%%%%%
% % 5.3
% %%%%%%%%%%%%%%%%%%%%%%%%%%%%%%%%%%%%%%%%%%%%%%%%%%
\subsection{Effect on Stellar Mass, Radius, Age, and Density}
\begin{figure*}
    \centering
    \includegraphics[width=1.0\linewidth]{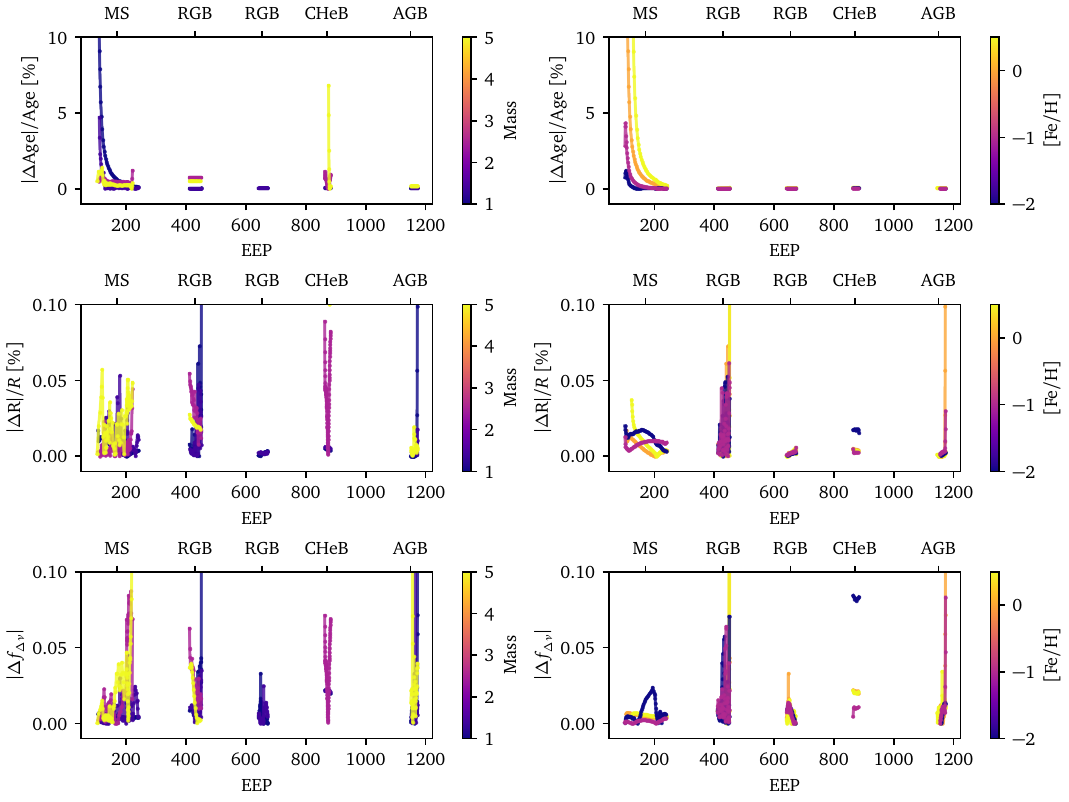}
    \caption{Differences in stellar parameters (Age, $R$, and \fDnu{}, from top row to bottom row, respectively) when evaluated against the same oscillation frequency for the two tracks with different \dmesh{}. The left columns show tracks with varying mass, and the right columns show those with varying metallicity.}
    \label{fig:radius_age}
\end{figure*}

We now estimate the influence of \dmesh{} on individual mode frequencies and the associated impact on deriving classical stellar parameters, including mass, radius and age. 
Additionally, we examine the model-based \fDnu{} factor, which is used to relate the \Dnu{} to the stellar mean density through \citep{Sharma2016}
\begin{equation}
    \frac{\Dnu{}}{135.1~{\mu\text{Hz}}} = \fDnu{} \left(\frac{\rho}{\rho_\odot}\right)^{0.5}.
\end{equation}

For this exercise, we chose a mode near the test frequency appropriate for each case (see Tables \ref{table:solar-like}, \ref{table:dSct} and \ref{table:Miras}) and determined interpolated values for radius, age, and \fDnu{} as functions of that mode frequency for two evolutionary tracks with different \dmesh{} values. Because mode frequencies are not strictly monotonic with evolution, we focus on only a few segments of the tracks where monotonicity is satisfied. We then computed differences in radius, age, and \fDnu{} at the same frequency. Figure~\ref{fig:radius_age} presents the fractional differences in radius and age, along with the absolute differences in \fDnu{}, as a function of evolutionary stage.

Asteroseismic constraints can achieve uncertainties of approximately 2\% in radius, 4\% in mass, and 10\% in age for red giants \citep{Montalban2021,Litd2022,Pinsonneault2024} and cool main-sequence dwarfs \citep{SilvaAguirre2015,SilvaAguirre2017} in optimal conditions. These uncertainties can be even lower for $\delta$ Scuti stars \citep{Murphy2021}. 
Determining these parameters for AGB stars is difficult, but Miras with two modes identified have been constrained to within 8\% in mass and 20\% in age \citep{Molnar2019}.

Figure \ref{fig:radius_age} therefore shows that,
for early main-sequence dwarfs, age uncertainties induced by \dmesh{} appear to approach the level of observational uncertainty. As stars evolve, their total ages increase while fractional age uncertainties decrease, reducing the relative impact of \dmesh{} on age estimation in later stages of evolution. Given the approximate scaling of stellar mass with age via $M \sim \tau^{-1/4}$, the uncertainties in mass induced by \dmesh{} are unlikely to exceed the typical 4\% observational uncertainties. Similarly, radius uncertainties fluctuate at a level of approximately 0.02\%, well below observational thresholds of 2\% but not entirely negligible. 

Fluctuations in \fDnu{} induced by \dmesh{}, however, are on the order of 0.02, which is comparable to those caused by other model physics \citep{Pinsonneault2024}. Hence, using the $\Delta\nu$ scaling relation with the model-based $\fDnu{}$ correction factors should be done carefully.

In the case of age, we find that resolution as a source of uncertainty is non-negligible, especially along the main sequence. A number of studies have examined the impact of modeling uncertainties---defined in various ways---on age determinations \citep[e.g.][]{Tayar2022,Joyce2023,YingChaboyer2023}.
The conclusion broadly shared by these investigations is that age error bars computed without modeling uncertainty considerations are heavily underestimated, and this study finds the same. 

%%%%%%%%%%%%%%%%%%%%%%%%%%%%%%%%%%%%%%%%%%%%%%%%%%
% % 5.4
% %%%%%%%%%%%%%%%%%%%%%%%%%%%%%%%%%%%%%%%%%%%%%%%%%
\subsection{Model Grid Emulators}
Several studies have aimed to emulate grids of stellar evolution calculations by approximating the relationships between stellar properties (e.g., mass, radius, age, oscillation frequencies) and model input parameters (e.g., initial mass, metallicity) using statistical models and machine learning techniques. These include methods such as normalizing flows \citep{Hon2024}, artificial neural networks \citep{Mombarg2021,Scutt2023}, Gaussian Processes \citep{Litd2022-gp}, and random forest \citep{Bellinger2016}. These approaches enable efficient interpolation across the complex and high-dimensional parameter space of model grids, producing smooth posteriors and significantly reducing computational costs (after the models are trained), enabling rapid parameter inference for large numbers of targets.

However, relatively few efforts have focused on achieving emulator precision significantly below observational uncertainties. One of the primary barriers to improving emulator precision is formally characterizing the numerical uncertainties inherent in the modeling process. This includes, but is not limited to, the effects of \dmesh{} examined in this study.

\section{Best Practices}
\label{sec:best_practices}
We have shown that inadequate structural resolution can be highly problematic for asteroseismic analyses, parameter determinations, and evolutionary inferences based on $p$-modes. However, this source of uncertainty can be
mitigated significantly (though not eliminated entirely) by adhering to a few key tenets of numerical modeling best practices:
\begin{enumerate}
    \item Choose parameter settings thoughtfully. 

    \item Perform convergence tests. 

    \item Balance resolution against run time. 

\end{enumerate}

To the first of these points, the most obvious lesson from the present analysis is that the default structural resolution in \texttt{MESA}, \texttt{delta\_mesh\_coeff}~$=1.0$, is not adequate for many of the asteroseismic applications discussed in this paper. As a general rule, one should never assume that the default parameter values provided in tools like \texttt{MESA} and \texttt{GYRE} are optimized for science problems. More often, these settings are optimized to run models quickly for teaching or training purposes. As such, it is the user's responsibility to choose resolution values---and all other parameter settings---that are appropriate for their problem.

Per point two, convergence tests, or resolution studies, are a necessary part of modeling science. Results that are presented without demonstrating that the inferences and/or conclusions drawn from models are robust against resolution changes should not be trusted. Performing convergence tests also means making the results of those convergence tests available, whether via inclusion in an appendix or by publicly hosting the analysis code on a platform like GitHub.

Lastly, convergence does not have an absolute standard, but rather is defined relative to the precision needed for the result. In this sense, it is no better to set the structural resolution arbitrarily high than it is to inherit default settings blindly: things like structural and temporal resolution must always be balanced against run time (with higher-resolution models taking longer to run) and strain on numerical solvers, which often struggle more as resolution increases.  
A rule-of-thumb indicator that results are converged is if the answer does not change \textit{within some reasonable tolerance} when the resolution is increased. The lowest resolution setting for which this is true is usually the optimal choice.
Examples of ``gold standard'' convergence testing can be found in, e.g.,  
\citet{Schootemeijer2019,Klencki2021} and \citet{Ziolkowska2024}.
Other studies that do not directly mention convergence testing but do
use appropriate, non-default values of \dmesh{} include \citet{Schmid2016,Buysschaert2018,Pedersen2018,Loi2020} and \citet{Johnston2024}. This is by no means a comprehensive list.

Even when abiding by all best practices in numerical modeling, there will inevitably be some discrepancy, and therefore uncertainty, caused by resolution choices, just as there is uncertainty introduced by selecting a particular set of physical assumptions or a particular modeling tool. To account for correlated mode uncertainties in asteroseismic model fitting, we recommend constructing the $\chi^2$ statistic using a covariance matrix, $\mathbf{C}$, as follows:
\begin{equation}
    \chi^2 = (\nu_{\rm obs}-\nu_{\rm mod})^{\mathbf{T}} \mathbf{C}^{-1} (\nu_{\rm obs}-\nu_{\rm mod}),
\end{equation}
where $\nu_{\rm obs}$ and $\nu_{\rm mod}$ represent the observed and modeled frequencies, respectively.

In this formulation, the covariance matrix, $\mathbf{C}$, accounts for the fact that mode uncertainties are not independent. 
We propose that all elements of $\mathbf{C}$ be populated using $(\dm{})^2$. This ensures the inclusion of numerical uncertainties alongside observational noise in the fitting process. 

Although we did not explicitly show the resolution-induced uncertainty for mode of different radial order, our calculations indicate that these uncertainties are largely correlated. This behavior is expected because all frequencies are derived from the same structural profile. Any resolution error in the structure will propagate to all frequencies, though the magnitude of the effect varies depending on each mode’s sensitivity to the probed region. In other contexts, \citet{Aerts2018} and \citet{Litd2022-gp} similarly found that modeling uncertainties in oscillation frequencies are correlated.

Incorporating a term like this may partially alleviate the need to manually down-weight seismic $\chi^2$ constraints \citep[e.g.][]{Cunha2021} as a means of compensating for ultra-small observational uncertainties on individual $p$-mode measurements.

\section{Summary}
\label{sec:conclusions}
In this study, we have quantified the impact of a factor-of-ten variation in the parameter \dmesh{} on individual $p$-mode frequencies for solar-like, upper main-sequence, and Mira oscillators as well as a number of asteroseismic and fundamental stellar parameters derived from these modes.
Using \texttt{MESA}'s default mesh size as the low-resolution reference point, we compared quantities determined at this resolution to those determined at a resolution 10x greater, introducing the quantities \dm{} and \dPm{} as measures of the discrepancy in frequency and period, respectively. These quantities then served as indicators of the uncertainty due to structural resolution for our chosen case of \texttt{delta\_mesh\_coeff}= 1 compared to \texttt{delta\_mesh\_coeff}= 0.1.
Our analysis required the development of a modified Equivalent Evolutionary Point (EEP) framework that can be readily adapted to study other forms of modeling uncertainty.

While solar-like oscillators typically had fractional uncertainties (\dm{}/\numax{}) at or below 1\% of the test frequency, fractional uncertainties in Miras (\dPm{}/$P_\text{test}$) were as large as 20\%.
We found that structural resolution uncertainty was largest in models with higher masses, higher metallicities, and occupying later evolutionary stages. This is consistent with expectations given the more complicated interior physics of stars of these types. 

We compared mesh resolution uncertainties to synthetic noise, $\sigma_\text{obs}$, computed for six TESS observing scenarios. In almost all cases uncertainty, \dm{} (\dPm{}) exceeded typical observational uncertainties, and often by many orders of magnitude. We computed the ratio of \dm{} (\dPm{}) to $\sigma_\text{obs}$ for the worst-case observing scenario (T = 1 month, $\sigma_\text{noise} = 197$ ppm/hr) across all evolutionary tracks and found that the ratio was highest in roughly the same regions where absolute structural resolution uncertainty was largest.

When investigating the impact of resolution uncertainty on quantities influenced by or derived from $p$-modes, we found that the location and morphology of the RGB bump and red clump were impacted substantially. Of the global seismic parameters we studied, \DPi{1} was the most heavily affected.

When examining the influence of resolution uncertainty on fundamental stellar parameters, young main-sequence stars fared the worst: ages were impacted at the 10\% level. Across models of nearly every evolutionary stage, \fDnu{} fluctuated at the 2\% level in response to changes in \dmesh{}, which is at the same level as uncertainties induced by changing model physics.

Though we focus on $p$-mode asteroseismology and \texttt{MESA}'s default structural resolution in this analysis, our results clearly demonstrate the importance of incorporating resolution-based uncertainties into all asteroseismic applications and the dangers of using an inappropriate mesh in seismic modeling in general. Currently, very few studies do incorporate any measure of mesh-related uncertainty (see, however, recent advances in methods for characterizing numerical error in stellar oscillation codes: \citealt{Townsend2025}). We strongly encourage a new standard in the field: when using models for asteroseismic analysis, take care to set \dmesh{} (and related numerical parameters) appropriately, perform and publish convergence tests, and incorporate a mesh resolution term when quoting uncertainties.

\section*{Acknowledgements}
%\begin{acknowledgements}
Y.\ Li and M.\ Joyce contributed equally to this manuscript and may both refer to this as a first-author publication.
The authors thank Jamie Tayar, Daniel Huber and L\'aszl\'o Moln\'ar for fruitful discussions and valuable suggestions.
Y.\ Li acknowledges support from Beatrice Watson Parrent Fellowship.
M.J. gratefully acknowledges funding of MATISSE: \textit{Measuring Ages Through Isochrones, Seismology, and Stellar Evolution}, project number 101038062, awarded through the European Commission's Widening Fellowship. This project has received funding from the European Union's Horizon 2020 research and innovation programme.
%\end{acknowledgements}
%%%%%%%%%%%%%%%%%%%%%%%%%%%%%%%%%%%%%%%%%%%%%%%%%%%%%%%%

\bibliography{astero_convergence.bib}
\bibliographystyle{mnras}

\appendix

\section{Primary EEP} 

\begin{figure*}%[htp] 
\includegraphics[width=\textwidth]{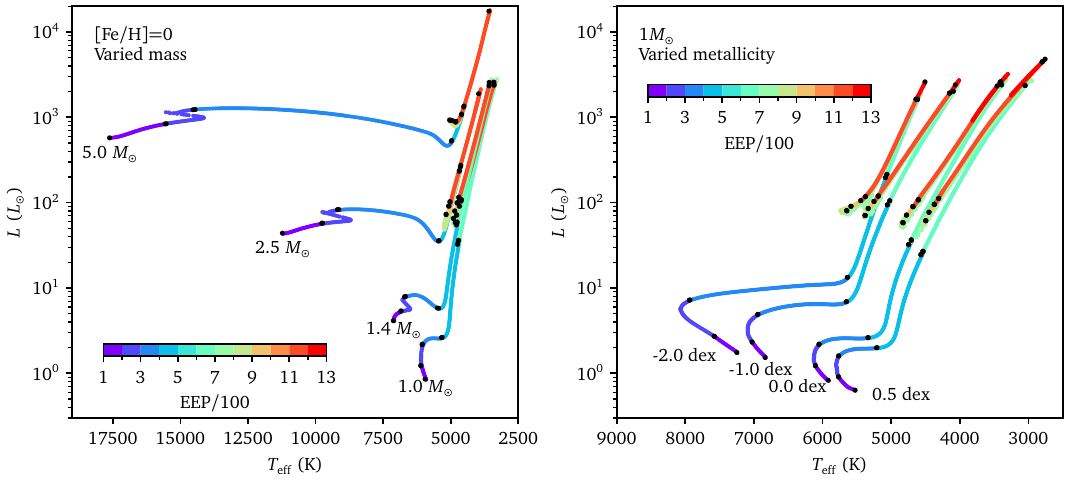}
\caption{Evolutionary tracks used in this studied color-coded by the primary EEP values. The black dots show the critical points that separate primary EEPs.}
\label{fig:primary_eep}
\end{figure*}

Figure~\ref{fig:primary_eep} shows the stellar models on the H--R diagram, color-coded by their primary EEP.

\section{Convergence Tests}
\begin{figure}%[htp] 
\centering\includegraphics[width=0.5\textwidth]{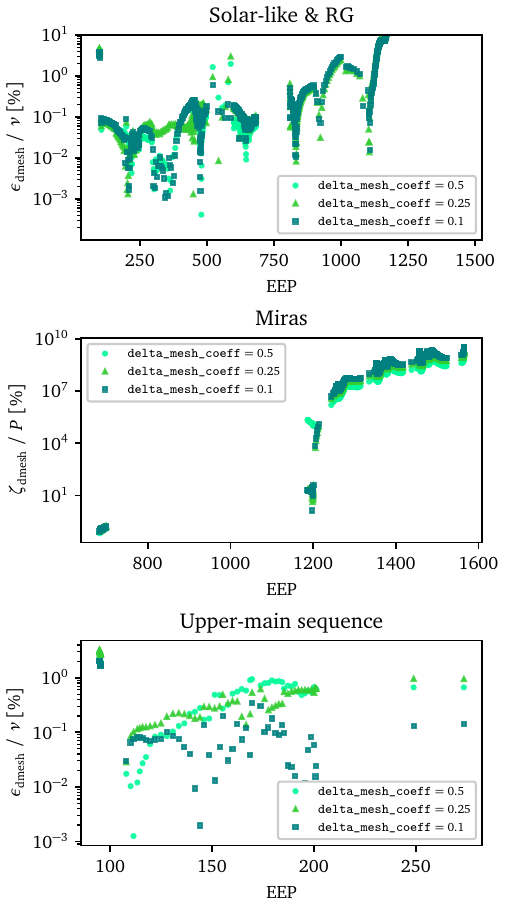}
\caption{Results of resolution tests, showing differences in frequency or period, \dm{} or \dPm{} (see Eqs.~\ref{eqn:dm} and \ref{eqn:dPm}), calculated using alternative \dmesh{} values to compare with the default \dmesh{}=1.0.}
\label{fig:test_dmesh}
\end{figure}

We selected a 1.0$M_\odot$, solar-metallicity evolutionary track to represent solar-like pulsators and Miras, and a 1.4$M_\odot$, solar-metallicity track to represent upper main-sequence pulsators. For each case, we computed four mesh resolutions using \texttt{delta\_mesh\_coeff} = 1.0, 0.5, 0.25, and 0.1. The figure below shows $\dm$ (or $\dPm$ for Miras) between the finer resolutions and the base case with \texttt{delta\_mesh\_coeff} = 1.0.
We find that, in general, the differences do not increase substantially with finer resolution, except for the Mira case. Based on these results, we consider \texttt{delta\_mesh\_coeff} = 0.1 to be a reasonable choice for higher resolution. We do not explore values lower than 0.1 in order to maintain a practical balance between resolution and computational cost.
Results from resolution tests are shown in Figure \ref{fig:test_dmesh}.

\end{document}

%% file: tables/table1_inlist_summary.tex
\begin{table*}[ht]
\centering
\begin{tabular}{| l | l | l |}
\toprule
\multicolumn{1}{|c}{\textbf{Physics Component}} &
\multicolumn{1}{|c|}{\textbf{Description}} &
\multicolumn{1}{c|}{\textbf{Reference}} \\
%------------------------------------------
\hline
Abundances & GS98 & \citet{gs98} \\ \hline
Nuclear reaction network & \texttt{`pp\_and\_cno\_extras\_o18\_ne22.net'}  &  \\ \
& includes all isotopes necessary for hydrogen-burning chains & \\ 
\hline
Equation of State & \texttt{Skye} EOS, \texttt{Free} EOS & \citet{MESAVI} \\ 
\hline 
Opacities  & GS98 for default &  \\
& with GS98 compatible Low-T, and CO opacities & \\ 
\hline
Mixing Length Theory & \texttt{`Henyey'}; appropriate for low optical depth & \citet{henyey1965}  \\ 
$\alpha_\text{MLT}$  & fixed to 2.1 & \\ 
\hline
Atmospheric boundary conditions & Eddington $T-\tau$  & \citet{eddington1926}  \\
 & \texttt{atm\_T\_tau\_opacity}=`varying' & \\ 
\hline
Heavy element diffusion & None & \\ 
\hline
%5 species; diffused as $^{56}$Fe & Thoul 1994
Treatment of convective boundaries & predictive mixing algorithm & \citet{mesa2018} \\ 
\hline
Convective overshoot & `step' for all regions & \\
 & $f = 0.02$ H$_\text{P}$ & \\
 & $f_0 = 0.002$ H$_\text{P}$ & \\ 
\hline
\texttt{MESA} numerical parameters & \texttt{varcontrol\_target} = \texttt{1e-4} & \citet{MESAVI} \\
& \texttt{time\_delta\_coeff} = \texttt{1.0}& \\
& \texttt{use\_gold\_tolerances} = \texttt{.true.}& \\
& \texttt{use\_gold2\_tolerances} = \texttt{.true.}&  \\
\hline
\texttt{GYRE} parameters 
& \texttt{diff\_scheme} = \texttt{`MAGNUS\_GL6'} & \citet{GYRE} \\
& \texttt{n\_iter\_max} = \texttt{50} & \\
& \texttt{inner\_bound} = \texttt{`REGULAR'} & \\
& \texttt{outer\_bound} = \texttt{`JCD'} & \\
\hline
\end{tabular}
\caption{Summary of physical assumptions used in \texttt{MESA} and \texttt{GYRE} models.}
\label{table:inlist_summary}
\end{table*} 

%% file: tables/table2_overview_model_classes.tex
\begin{table*}[ht]
\centering
\begin{tabular}{| c | c | c | c | c | c | c | c | c | c |} %%% 10 columns
\toprule
\multicolumn{1}{|c|}{\textbf{Mass}} &
\multicolumn{1}{c|}{\textbf{Z}} &
\multicolumn{1}{c|}{\textbf{[Fe/H]}} &
\multicolumn{1}{c|}{\textbf{Ev Phase}} &
\multicolumn{1}{c|}{\textbf{EEP}} &
\multicolumn{1}{c|}{\textbf{Age}} &

\multicolumn{1}{c|}{\textbf{Teff}} &
\multicolumn{1}{c|}{\textbf{$\log$L}} &
\multicolumn{1}{c|}{\textbf{$\log$g}} &
\multicolumn{1}{c|}{\textbf{see table...}}

%\multicolumn{1}{c|}{\textbf{$\nu_\text{max}$}} &
%\multicolumn{1}{c|}{\textbf{$\nu_\text{test}$}} &
%\multicolumn{1}{c|}{\textbf{$n$}} &
%\multicolumn{1}{c|}{\textbf{$\dm$}} &
%\multicolumn{1}{c|}{\textbf{\%$\dm$}} 
\\
\hline
$M_\odot$ & 
&
dex & 
&
&
Gyr &
K &
L$_\odot$ &
  &
\\ \hline
\multicolumn{10}{|l|}{ \textbf{Model Class 1: Solar-like} }\\\hline
1.0 & 0.0162 & 0.0 & MS & 187 & 3.870 & 6091 & 0.070 & 4.461 & solar-like; Table \ref{table:solar-like} \\
1.0 & 0.0162 & 0.0 & subgiant & 303 & 8.750 & 6040 & 0.348 & 4.169 & solar-like; Table \ref{table:solar-like} \\
1.0 & 0.0162 & 0.0 & early RGB & 410 & 9.791 & 5256 & 0.433 & 3.843 & solar-like; Table \ref{table:solar-like} \\
1.0 & 0.0162 & 0.0 & late RGB & 665 & 10.404 & 3802 & 2.886 & 0.827 & solar-like; Table \ref{table:solar-like} \\
1.0 & 0.0162 & 0.0 & CHeB & 850 & 10.450 & 4862 & 1.765 & 2.375 & solar-like; Table \ref{table:solar-like} \\
1.0 & 0.0162 & 0.0 & AGB & 1198 & 10.503 & 3488 & 3.334 & 0.229 & Mira variables; Table \ref{table:Miras} \\
\hline
\multicolumn{10}{|l|}{ \textbf{Model Class 2: metal-rich, solar mass} }\\\hline
1.0 & 0.0447 & +0.5 & MS & 187 & 3.377 & 5754 & -0.059 & 4.492 & solar-like; Table \ref{table:solar-like} \\
1.0 & 0.0447 & +0.5 & subgiant & 303 & 9.415 & 5757 & 0.212 & 4.222 & solar-like; Table \ref{table:solar-like} \\
1.0 & 0.0447 & +0.5 & early RGB & 410 & 10.681 & 5134 & 0.310 & 3.925 & solar-like; Table \ref{table:solar-like} \\
1.0 & 0.0447 & +0.5 & late RGB & 665 & 11.330 & 3543 & 2.854 & 0.736 & solar-like; Table \ref{table:solar-like} \\
1.0 & 0.0447 & +0.5 & CHeB & 850 & 11.378 & 4546 & 1.771 & 2.252 & solar-like; Table \ref{table:solar-like} \\
1.0 & 0.0447 & +0.5 & AGB & 1198 & 11.438 & 2870 & 3.591 & -0.366 & Mira variables; Table \ref{table:Miras} \\
\hline
\multicolumn{10}{|l|}{ \textbf{Model Class 3: metal-poor, solar mass} }\\\hline
1.0 & 0.0017 & -1.0 & MS & 187 & 2.564 & 7010 & 0.347 & 4.429 & Upper main sequence; Table \ref{table:dSct} \\
1.0 & 0.0017 & -1.0 & subgiant & 303 & 5.507 & 6912 & 0.703 & 4.048 & Upper main sequence; Table \ref{table:dSct} \\
1.0 & 0.0017 & -1.0 & early RGB & 410 & 6.019 & 5580 & 0.885 & 3.494 & solar-like; Table \ref{table:solar-like} \\
1.0 & 0.0017 & -1.0 & late RGB & 665 & 6.289 & 4394 & 2.926 & 1.039 & solar-like; Table \ref{table:solar-like} \\
1.0 & 0.0017 & -1.0 & CHeB & 850 & 6.330 & 5406 & 1.869 & 2.455 & solar-like; Table \ref{table:solar-like} \\
1.0 & 0.0017 & -1.0 & AGB & 1198 & 6.375 & 4218 & 3.209 & 0.684 & Mira variables; Table \ref{table:Miras} \\
\hline
\multicolumn{10}{|l|}{ \textbf{Model Class 4: very metal-poor, solar mass} }\\\hline
1.0 & 0.0002 & -2.0 & MS & 187 & 2.357 & 7539 & 0.412 & 4.490 & Upper main sequence; Table \ref{table:dSct} \\
1.0 & 0.0002 & -2.0 & subgiant & 303 & 5.160 & 7901 & 0.869 & 4.114 & Upper main sequence; Table \ref{table:dSct} \\
1.0 & 0.0002 & -2.0 & early RGB & 410 & 5.496 & 5575 & 1.194 & 3.184 & solar-like; Table \ref{table:solar-like} \\
1.0 & 0.0002 & -2.0 & late RGB & 665 & 5.650 & 4765 & 2.933 & 1.172 & solar-like; Table \ref{table:solar-like} \\
1.0 & 0.0002 & -2.0 & CHeB & 850 & 5.687 & 5741 & 1.910 & 2.519 & solar-like; Table \ref{table:solar-like} \\
1.0 & 0.0002 & -2.0 & AGB & 1198 & 5.729 & 4673 & 3.156 & 0.916 & Mira variables; Table \ref{table:Miras} \\
\hline
\multicolumn{10}{|l|}{ \textbf{Model Class 5: solar metallicity, $1.4M_\odot$} }\\\hline
1.4 & 0.0162 & 0.0 & MS & 187 & 1.457 & 6926 & 0.720 & 4.181 & Upper main sequence; Table \ref{table:dSct} \\
1.4 & 0.0162 & 0.0 & subgiant & 303 & 2.460 & 6657 & 0.904 & 3.929 & solar-like; Table \ref{table:solar-like} \\
1.4 & 0.0162 & 0.0 & early RGB & 410 & 2.788 & 5384 & 0.767 & 3.696 & solar-like; Table \ref{table:solar-like} \\
1.4 & 0.0162 & 0.0 & late RGB & 665 & 3.125 & 3927 & 2.925 & 0.990 & solar-like; Table \ref{table:solar-like} \\
1.4 & 0.0162 & 0.0 & CHeB & 850 & 3.169 & 4928 & 1.817 & 2.493 & solar-like; Table \ref{table:solar-like} \\
1.4 & 0.0162 & 0.0 & AGB & 1198 & 3.222 & 3592 & 3.392 & 0.368 & Mira variables; Table \ref{table:Miras} \\
\hline
\multicolumn{10}{|l|}{ \textbf{Model Class 6: solar metallicity, $2.5M_\odot$} }\\\hline
2.5 & 0.0162 & 0.0 & MS & 187 & 0.319 & 9902 & 1.751 & 4.022 & Upper main sequence; Table \ref{table:dSct} \\
2.5 & 0.0162 & 0.0 & subgiant & 303 & 0.470 & 9087 & 1.921 & 3.703 & Upper main sequence; Table \ref{table:dSct} \\
2.5 & 0.0162 & 0.0 & early RGB & 410 & 0.484 & 5326 & 1.604 & 3.092 & solar-like; Table \ref{table:solar-like} \\
2.5 & 0.0162 & 0.0 & late RGB & 495 & 0.491 & 4723 & 2.321 & 2.167 & solar-like; Table \ref{table:solar-like} \\
2.5 & 0.0162 & 0.0 & CHeB & 850 & 0.497 & 5090 & 1.877 & 2.741 & solar-like; Table \ref{table:solar-like} \\
2.5 & 0.0162 & 0.0 & AGB & 1198 & 0.684 & 3944 & 3.305 & 0.870 & Mira variables; Table \ref{table:Miras} \\
\hline
\multicolumn{10}{|l|}{ \textbf{Model Class 7: solar metallicity, $5.0M_\odot$} }\\\hline
5.0 & 0.0162 & 0.0 & MS & 187 & 0.057 & 15785 & 2.909 & 3.976 & Upper main sequence; Table \ref{table:dSct} \\
5.0 & 0.0162 & 0.0 & subgiant & 303 & 0.081 & 14323 & 3.094 & 3.622 & Upper main sequence; Table \ref{table:dSct} \\
5.0 & 0.0162 & 0.0 & early RGB & 410 & 0.082 & 4875 & 2.786 & 2.058 & solar-like; Table \ref{table:solar-like} \\
5.0 & 0.0162 & 0.0 & late RGB & 495 & 0.082 & 4608 & 3.026 & 1.720 & solar-like; Table \ref{table:solar-like} \\
5.0 & 0.0162 & 0.0 & CHeB & 850 & 0.087 & 4662 & 2.998 & 1.768 & solar-like; Table \ref{table:solar-like} \\
5.0 & 0.0162 & 0.0 & AGB & 1198 & 0.100 & 3604 & 4.224 & 0.095 & Mira variables; Table \ref{table:Miras} \\

\hline
\end{tabular}
\caption{Overview of model classes and summary of fundamental parameters.}
\label{table:overview}
\end{table*}

%% file: tables/table3_solar-like.tex
\begin{table*}[ht]
\centering
\hspace*{-90pt}%
\begin{tabular}{| c | c | c | c | c | c | c | c | c | c | c | c |} %%% 12 columns
\toprule
\multicolumn{1}{|c|}{\textbf{Mass}} &
\multicolumn{1}{c|}{\textbf{Z}} &
\multicolumn{1}{c|}{\textbf{[Fe/H]}} &
\multicolumn{1}{c|}{\textbf{Ev Phase}} &
% \multicolumn{1}{c|}{\textbf{Star Type}} &
\multicolumn{1}{c|}{\textbf{EEP}} &
\multicolumn{1}{c|}{\textbf{Age}} &

\multicolumn{1}{c|}{\textbf{$n$}} &
\multicolumn{1}{c|}{\textbf{$\nu_\text{max}$}} &

\multicolumn{1}{c|}{\textbf{$\dm$}} &
\multicolumn{1}{c|}{\textbf{\%$\dm$}} & 
\multicolumn{1}{c|}{\textbf{$n_\text{zones}$}} & 
\multicolumn{1}{c|}{\textbf{$n_\text{zones}$}} 
\\
\hline
$M_\odot$ & 
&
dex & 
&
&
Gyr &

$\ell, m  = 0$ &
$\mu$Hz &

$\mu$Hz & %$\Delta$(\texttt{dmesh}=1.0, \texttt{dmesh}=0.1) &
$\dm$/$\nu_\text{max}$ &
\texttt{[dmesh=1.0]} &
\texttt{[dmesh=0.1]} \\ \hline
\multicolumn{12}{|l|}{ \textbf{Model Class 1: Solar-like} }\\\hline
1.0 & 0.0162 & 0.0 & MS & 187 & 3.870 & 21,22 & 3173.8 & 1.0355 & 0.0326 & 759 & 7261 \\
1.0 & 0.0162 & 0.0 & subgiant & 303 & 8.749 & 17,18 & 1626.0 & 0.0655 & 0.0040 & 852 & 8134 \\
1.0 & 0.0162 & 0.0 & early RGB & 410 & 9.791 & 15,16 & 821.6 & 0.8458 & 0.1030 & 1053 & 10032 \\
1.0 & 0.0162 & 0.0 & late RGB & 665 & 10.404 & 3,4 & 0.9 & 0.0006 & 0.0610 & 1621 & 15207 \\
1.0 & 0.0162 & 0.0 & CHeB & 850 & 10.450 & 6,7 & 29.1 & 0.0843 & 0.2891 & 1338 & 12603 \\
\hline
\multicolumn{12}{|l|}{ \textbf{Model Class 2: metal-rich, solar mass} }\\\hline
1.0 & 0.0447 & +0.5 & MS & 187 & 3.377 & 22,23 & 3501.9 & 1.3584 & 0.0388 & 771 & 7313 \\
1.0 & 0.0447 & +0.5 & subgiant & 303 & 9.415 & 18,19 & 1879.8 & 1.3039 & 0.0692 & 857 & 8108 \\
1.0 & 0.0447 & +0.5 & early RGB & 410 & 10.681 & 16,17 & 1005.3 & 5.6531 & 0.5624 & 1045 & 9947 \\
1.0 & 0.0447 & +0.5 & late RGB & 665 & 11.330 & 3,4 & 0.8 & 0.0028 & 0.3605 & 1601 & 15175 \\
1.0 & 0.0447 & +0.5 & CHeB & 850 & 11.378 & 6,7 & 22.7 & 0.0387 & 0.1706 & 1413 & 13406 \\
\hline
\multicolumn{12}{|l|}{ \textbf{Model Class 3: metal-poor, solar mass} }\\\hline
1.0 & 0.0017 & -1.0 & early RGB & 410 & 6.019 & 12,13 & 357.7 & 0.2409 & 0.0674 & 1140 & 10785 \\
1.0 & 0.0017 & -1.0 & late RGB & 665 & 6.289 & 3,4 & 1.4 & 0.0011 & 0.0814 & 1598 & 15746 \\
1.0 & 0.0017 & -1.0 & CHeB & 850 & 6.330 & 6,7 & 33.2 & 0.0121 & 0.0365 & 1281 & 12061 \\
\hline
\multicolumn{12}{|l|}{ \textbf{Model Class 4: very metal-poor, solar mass} }\\\hline
1.0 & 0.0002 & -2.0 & early RGB & 410 & 5.496 & 10,11 & 175.1 & 0.5487 & 0.3135 & 1193 & 11501 \\
1.0 & 0.0002 & -2.0 & late RGB & 665 & 5.650 & 3,4 & 1.8 & 0.0081 & 0.4397 & 1616 & 15813 \\
1.0 & 0.0002 & -2.0 & CHeB & 850 & 5.687 & 6,7 & 37.3 & 0.0551 & 0.1475 & 1293 & 11972 \\
\hline
\multicolumn{12}{|l|}{ \textbf{Model Class 5: solar metallicity, $1.4M_\odot$} }\\\hline
1.4 & 0.0162 & 0.0 & subgiant & 303 & 2.460 & 16,17 & 890.2 & 0.4397 & 0.0493 & 968 & 9001 \\
1.4 & 0.0162 & 0.0 & early RGB & 410 & 2.788 & 15,16 & 579.5 & 1.7503 & 0.3022 & 1126 & 10892 \\
1.4 & 0.0162 & 0.0 & late RGB & 665 & 3.125 & 3,4 & 1.3 & 0.0072 & 0.5417 & 1629 & 15700 \\
1.4 & 0.0162 & 0.0 & CHeB & 850 & 3.169 & 7,8 & 37.9 & 0.0147 & 0.0388 & 1346 & 12612 \\
\hline
\multicolumn{12}{|l|}{ \textbf{Model Class 6: solar metallicity, $2.5M_\odot$} }\\\hline
2.5 & 0.0162 & 0.0 & early RGB & 410 & 0.484 & 12,13 & 145.0 & 1.6072 & 1.1081 & 1176 & 12901 \\
2.5 & 0.0162 & 0.0 & late RGB & 495 & 0.491 & 7,8 & 18.3 & 0.1112 & 0.6067 & 1471 & 14236 \\
2.5 & 0.0162 & 0.0 & CHeB & 850 & 0.502 & 10,11 & 99.9 & 1.5029 & 1.4983 & 1437 & 13867 \\
\hline
\multicolumn{12}{|l|}{ \textbf{Model Class 7: solar metallicity, $5.0M_\odot$} }\\\hline
5.0 & 0.0162 & 0.0 & early RGB & 410 & 0.082 & 8,9 & 14.0 & 0.2220 & 1.5849 & 1256 & 13837 \\
5.0 & 0.0162 & 0.0 & late RGB & 495 & 0.082 & 7,8 & 6.6 & 0.1740 & 2.6303 & 1373 & 14743 \\
5.0 & 0.0162 & 0.0 & CHeB & 850 & 0.087 & 7,8 & 7.3 & 0.0605 & 0.8239 & 1525 & 13984 \\

\hline
\end{tabular}
\caption{Frequency discrepancies due to structural resolution, $\dm$, calculated for solar-like oscillators. $\dm$ is evaluated at $\nu_\text{max}$. The mode with radial order $n$ and spherical degree and azimuthal orders $\ell,m$ = 0 nearest to $\nu_\text{max}$ is shown. The number of mesh points $n_\text{zones}$ is also shown.}
\label{table:solar-like}
\end{table*}

%% file: tables/table4_dSct-like.tex
\begin{table*}[ht]
\centering
\hspace*{-90pt}%
\begin{tabular}{| c | c | c | c | c | c | c | c | c | c | c | c |} %%% 12 columns
\toprule
\multicolumn{1}{|c|}{\textbf{Mass}} &
\multicolumn{1}{c|}{\textbf{Z}} &
\multicolumn{1}{c|}{\textbf{[Fe/H]}} &
\multicolumn{1}{c|}{\textbf{Ev Phase}} &
\multicolumn{1}{c|}{\textbf{EEP}} &
\multicolumn{1}{c|}{\textbf{Age}} &

\multicolumn{1}{c|}{\textbf{$n$}} &
\multicolumn{1}{c|}{\textbf{$\nu_\text{test}$}} &

\multicolumn{1}{c|}{\textbf{$\dm$}} &
\multicolumn{1}{c|}{\textbf{\%$\dm$}} &
\multicolumn{1}{c|}{\textbf{$n_\text{zones}$}} & 
\multicolumn{1}{c|}{\textbf{$n_\text{zones}$}} 
\\
\hline
$M_\odot$ & 
&
dex & 
&
&
Myr &

&
cyc/day &
cyc/day & %$\Delta$(\texttt{dmesh}=1.0, \texttt{dmesh}=0.1) &
$\dm$/$\nu_\text{test}$ &
\texttt{[dmesh=1.0]} &
\texttt{[dmesh=0.1]} \\ \hline
\multicolumn{12}{|l|}{ \textbf{Model Class 3: metal-poor, solar mass} }\\\hline
1.0 & 0.0017 & -1.0 & MS & 187 & 2564.30 & 5 & 45.0 & 0.04 & 0.08 & 753 & 7247 \\
1.0 & 0.0017 & -1.0 & subgiant & 303 & 5506.43 & 5 & 24.9 & 0.02 & 0.10 & 860 & 8298 \\
\hline
\multicolumn{12}{|l|}{ \textbf{Model Class 4: very metal-poor, solar mass} }\\\hline
1.0 & 0.0002 & -2.0 & MS & 187 & 2357.09 & 5 & 51.0 & 0.08 & 0.16 & 786 & 7391 \\
1.0 & 0.0002 & -2.0 & subgiant & 303 & 5159.71 & 5 & 28.4 & 0.03 & 0.09 & 934 & 9003 \\
\hline
\multicolumn{12}{|l|}{ \textbf{Model Class 5: solar metallicity, $1.4M_\odot$} }\\\hline
1.4 & 0.0162 & 0.0 & MS & 187 & 1456.72 & 5 & 28.2 & 0.01 & 0.03 & 800 & 7704 \\
1.4 & 0.0162 & 0.0 & subgiant & 303 & 2356.82 & 5 & 20.8 & 0.03 & 0.16 & 923 & 9001 \\
\hline
\multicolumn{12}{|l|}{ \textbf{Model Class 6: solar metallicity, $2.5M_\odot$} }\\\hline
2.5 & 0.0162 & 0.0 & MS & 187 & 319.15 & 5 & 18.9 & 0.11 & 0.59 & 936 & 10138 \\
2.5 & 0.0162 & 0.0 & subgiant & 303 & 470.02 & 5 & 11.0 & 0.18 & 1.66 & 1109 & 13492 \\
\hline
\multicolumn{12}{|l|}{ \textbf{Model Class 7: solar metallicity, $5.0M_\odot$} }\\\hline
5.0 & 0.0162 & 0.0 & MS & 187 & 56.79 & 5 & 14.2 & 0.05 & 0.35 & 895 & 10152 \\
5.0 & 0.0162 & 0.0 & subgiant & 303 & 80.77 & 5 & 2.7 & 0.07 & 2.72 & 1055 & 13736 \\

\hline
\end{tabular}
\caption{Frequency discrepancies due to structural resolution, $\dm$, calculated for upper main-sequence pulsators (including $\delta$ Scutis). $\dm$ is evaluated for $n=3$, $\ell=0$ modes. Note that age is shown in Myr instead of Gyr. The number of mesh points $n_\text{zones}$ is shown.} 
\label{table:dSct}
\end{table*}

%% file: tables/table5_Miras.tex
\begin{table*}[ht]
\hspace*{-90pt}%
\centering
\begin{tabular}{| c | c | c | c | c | c | c | c | c | c | c | c |} %%% 12 columns
\toprule
\multicolumn{1}{|c|}{\textbf{Mass}} &
\multicolumn{1}{c|}{\textbf{Z}} &
\multicolumn{1}{c|}{\textbf{[Fe/H]}} &
\multicolumn{1}{c|}{\textbf{Ev Phase}} &
\multicolumn{1}{c|}{\textbf{EEP}} &
\multicolumn{1}{c|}{\textbf{Age}} &
\multicolumn{1}{c|}{\textbf{$n$}} &
\multicolumn{1}{c|}{\textbf{$P_\text{test}$}} &
\multicolumn{1}{c|}{\textbf{$\dPm$}} &
\multicolumn{1}{c|}{\textbf{\%$\dPm$}} & 
\multicolumn{1}{c|}{\textbf{$n_\text{zones}$}} & 
\multicolumn{1}{c|}{\textbf{$n_\text{zones}$}}
\\
\hline
$M_\odot$ & 
&
dex & 
&
&
Gyr &
$\ell, m =0$&
days &
&
$\dPm$/$P_\text{test}$ &
\texttt{[dmesh=1.0]} &
\texttt{[dmesh=0.1]} \\ \hline
\multicolumn{12}{|l|}{ \textbf{Model Class 1: Solar-like} }\\\hline
1.0 & 0.0162 & 0.0 & AGB & 1198 & 10.50 & 1 & 98.17 & 19.82 & 20.15 & 2010 & 20158 \\
\hline
\multicolumn{12}{|l|}{ \textbf{Model Class 2: metal-rich, solar mass} }\\\hline
1.0 & 0.0447 & +0.5 & AGB & 1198 & 11.44 & 1 & 389.74 & 8.04 & 2.07 & 2082 & 20290 \\
\hline
\multicolumn{12}{|l|}{ \textbf{Model Class 3: metal-poor, solar mass} }\\\hline
1.0 & 0.0017 & -1.0 & AGB & 1198 & 6.38 & 1 & 37.46 & 3.71 & 9.90 & 1900 & 18642 \\
\hline
\multicolumn{12}{|l|}{ \textbf{Model Class 4: very metal-poor, solar mass} }\\\hline
1.0 & 0.0002 & -2.0 & AGB & 1198 & 5.73 & 1 & 23.72 & 0.11 & 0.46 & 1917 & 18850 \\
\hline
\multicolumn{12}{|l|}{ \textbf{Model Class 5: solar metallicity, $1.4M_\odot$} }\\\hline
1.4 & 0.0162 & 0.0 & AGB & 1198 & 3.22 & 1 & 77.24 & 6.87 & 8.84 & 1987 & 20081 \\
\hline
\multicolumn{12}{|l|}{ \textbf{Model Class 6: solar metallicity, $2.5M_\odot$} }\\\hline
2.5 & 0.0162 & 0.0 & AGB & 1198 & 0.68 & 1 & 34.13 & 3.05 & 8.93 & 1876 & 19231 \\
\hline
\multicolumn{12}{|l|}{ \textbf{Model Class 7: solar metallicity, $5.0M_\odot$} }\\\hline
5.0 & 0.0162 & 0.0 & AGB & 1198 & 0.10 & 1 & 200.70 & 1.08 & 0.54 & 2344 & 22263 \\

\hline
\end{tabular}
\caption{Period discrepancies due to structural resolution, $\dPm$, calculated for Mira variables. $\dPm$ is evaluated at the fundamental mode frequency, $n=1,\ell=0, m=0$. The number of mesh points $n_\text{zones}$ is shown.} 
\label{table:Miras}
\end{table*}

%% file: tables/table6_compare_to_TESS_solar.tex
\begin{table*}[ht]
\centering
\begin{tabular}{| c | c | c | c | c | c| c | c| c| c|} %%% 10 columns
\toprule
\multicolumn{1}{|c|}{\textbf{mass}} &
\multicolumn{1}{c|}{\textbf{[Fe/H]}} &
\multicolumn{1}{c|}{\textbf{Phase}} &
\multicolumn{1}{c|}{\textbf{$\dm$}} &
\multicolumn{1}{c|}{\textbf{ $\dm$/$\sigma_\text{obs}$}} &
\multicolumn{1}{c|}{\textbf{$\dm$/$\sigma_\text{obs}$}} &
\multicolumn{1}{c|}{\textbf{$\dm$/$\sigma_\text{obs}$}} &
\multicolumn{1}{c|}{\textbf{$\dm$/$\sigma_\text{obs}$}} &
\multicolumn{1}{c|}{\textbf{$\dm$/$\sigma_\text{obs}$}} &
\multicolumn{1}{c|}{\textbf{$\dm$/$\sigma_\text{obs}$}} 
\\
\hline
& 
&
& 
$\mu$Hz &
T=4 yrs &
T=1 yrs &
T=0.083 yrs &
T=4 yrs &
T=1 yrs &
T=0.083 yrs \\ 
& 
&
& 
&
$\sigma_\text{noise}$=19.1 &
$\sigma_\text{noise}$=19.1 &
$\sigma_\text{noise}$=19.1 &
$\sigma_\text{noise}$=197 &
$\sigma_\text{noise}$=197 &
$\sigma_\text{noise}$=197 \\ 
\hline
\multicolumn{10}{|l|}{ \textbf{Model Class 1: Solar-like} }\\\hline
1.0 & 0.0 & MS & 1.035 & 48.5 & 24.2 & 6.4 & 4.7 & 2.4 & 0.6 \\
1.0 & 0.0 & subgiant & 0.065 & 5.8 & 2.9 & 0.8 & 0.6 & 0.3 & 0.1 \\
1.0 & 0.0 & early RGB & 0.846 & 192 & 95.6 & 22.4 & 18.6 & 9.3 & 2.2 \\
1.0 & 0.0 & late RGB & 0.001 & 1490 & 261 & 6.5 & 145 & 25.4 & 0.6 \\
1.0 & 0.0 & CHeB & 0.084 & 582 & 289 & 48.4 & 56.5 & 28.1 & 4.7 \\
\hline
\multicolumn{10}{|l|}{ \textbf{Model Class 2: metal-rich, solar mass} }\\\hline
1.0 & +0.5 & MS & 1.358 & 66.2 & 33.1 & 8.6 & 6.4 & 3.2 & 0.8 \\
1.0 & +0.5 & subgiant & 1.304 & 113 & 56.5 & 14.5 & 11.0 & 5.5 & 1.4 \\
1.0 & +0.5 & early RGB & 5.653 & 1112 & 555 & 128 & 108 & 53.9 & 12.4 \\
1.0 & +0.5 & late RGB & 0.003 & 8751 & 1174 & 28.2 & 849 & 114 & 2.7 \\
1.0 & +0.5 & CHeB & 0.039 & 770 & 370 & 30.3 & 74.8 & 36.0 & 2.9 \\
\hline
\multicolumn{10}{|l|}{ \textbf{Model Class 3: metal-poor, solar mass} }\\\hline
1.0 & -1.0 & early RGB & 0.241 & 31.5 & 15.7 & 4.4 & 3.1 & 1.5 & 0.4 \\
1.0 & -1.0 & late RGB & 0.001 & 685 & 305 & 15.3 & 66.5 & 29.6 & 1.5 \\
1.0 & -1.0 & CHeB & 0.012 & 19.7 & 9.8 & 2.7 & 1.9 & 1.0 & 0.3 \\
\hline
\multicolumn{10}{|l|}{ \textbf{Model Class 4: very metal-poor, solar mass} }\\\hline
1.0 & -2.0 & early RGB & 0.549 & 137 & 68.6 & 19.4 & 13.3 & 6.7 & 1.9 \\
1.0 & -2.0 & late RGB & 0.008 & 1672 & 824 & 103 & 162 & 80.0 & 10.0 \\
1.0 & -2.0 & CHeB & 0.055 & 37.4 & 18.7 & 5.3 & 3.6 & 1.8 & 0.5 \\
\hline
\multicolumn{10}{|l|}{ \textbf{Model Class 5: solar metallicity, $1.4M_\odot$} }\\\hline
1.4 & 0.0 & subgiant & 0.440 & 45.3 & 22.6 & 6.1 & 4.4 & 2.2 & 0.6 \\
1.4 & 0.0 & early RGB & 1.750 & 445 & 222 & 52.7 & 43.2 & 21.6 & 5.1 \\
1.4 & 0.0 & late RGB & 0.007 & 16777 & 3746 & 96.9 & 1629 & 364 & 9.4 \\
1.4 & 0.0 & CHeB & 0.015 & 57.9 & 28.8 & 5.4 & 5.6 & 2.8 & 0.5 \\
\hline
\multicolumn{10}{|l|}{ \textbf{Model Class 6: solar metallicity, $2.5M_\odot$} }\\\hline
2.5 & 0.0 & early RGB & 1.607 & 546 & 273 & 73.4 & 53.0 & 26.5 & 7.1 \\
2.5 & 0.0 & late RGB & 0.111 & 2604 & 1283 & 159 & 253 & 125 & 15.5 \\
2.5 & 0.0 & CHeB & 1.503 & 936 & 468 & 120 & 91.0 & 45.4 & 11.6 \\
\hline
\multicolumn{10}{|l|}{ \textbf{Model Class 7: solar metallicity, $5.0M_\odot$} }\\\hline
5.0 & 0.0 & early RGB & 0.222 & 16829 & 8360 & 1399 & 1635 & 812 & 136 \\
5.0 & 0.0 & late RGB & 0.174 & 78510 & 38073 & 3448 & 7626 & 3698 & 335 \\
5.0 & 0.0 & CHeB & 0.061 & 20616 & 10080 & 1045 & 2002 & 979 & 101 \\

\hline
\end{tabular}
\caption{Comparison of resolution-induced numerical vs observational uncertainties for solar-like ($p$-mode) oscillators.
The observational uncertainties are evaluated for six different combinations of observing strategies, defined by three observing durations ($T$ = 1 month, 1 year, and 4 years) and two noise levels ($\sigma_{\rm noise}$ = 19.1 ppm/hr and 197 ppm/hr), which corresponds to 5th and 10th magnitude TESS stars or 10th and 15th magnitude \Kepler{} stars. We estimate stellar oscillation amplitudes based on their stellar properties, including mass, luminosity, and \Teff{}.}
\label{table:TESS_solar}
\end{table*}

%% file: tables/table7_compare_to_TESS_dSct.tex
\begin{table*}[ht]
\centering
\begin{tabular}{| c | c | c | c | c | c| c | c| c| c|} %%% 10 columns
\toprule
\multicolumn{1}{|c|}{\textbf{mass}} &
\multicolumn{1}{c|}{\textbf{[Fe/H]}} &
\multicolumn{1}{c|}{\textbf{Phase}} &
\multicolumn{1}{c|}{\textbf{$\dm$}} &
\multicolumn{1}{c|}{\textbf{ $\dm$/$\sigma_\text{obs}$}} &
\multicolumn{1}{c|}{\textbf{$\dm$/$\sigma_\text{obs}$}} &
\multicolumn{1}{c|}{\textbf{$\dm$/$\sigma_\text{obs}$}} &
\multicolumn{1}{c|}{\textbf{$\dm$/$\sigma_\text{obs}$}} &
\multicolumn{1}{c|}{\textbf{$\dm$/$\sigma_\text{obs}$}} &
\multicolumn{1}{c|}{\textbf{$\dm$/$\sigma_\text{obs}$}} 
\\
\hline
& 
&
& 
cyc/day &
T=4 yrs &
T=1 yrs &
T=0.083 yrs &
T=4 yrs &
T=1 yrs &
T=0.083 yrs \\ 
& 
&
& 
&
$\sigma_\text{noise}$=19.1 &
$\sigma_\text{noise}$=19.1 &
$\sigma_\text{noise}$=19.1 &
$\sigma_\text{noise}$=197 &
$\sigma_\text{noise}$=197 &
$\sigma_\text{noise}$=197 \\ 
\hline
\multicolumn{10}{|l|}{ \textbf{Model Class 3: metal-poor, solar mass} }\\\hline
1.0 & -1.0 & MS & 0.04 & 9669 & 1209 & 28.9 & 940 & 118 & 2.8 \\
1.0 & -1.0 & subgiant & 0.02 & 6118 & 765 & 18.3 & 595 & 74.4 & 1.8 \\
\hline
\multicolumn{10}{|l|}{ \textbf{Model Class 4: very metal-poor, solar mass} }\\\hline
1.0 & -2.0 & MS & 0.08 & 21119 & 2640 & 63.1 & 2054 & 257 & 6.1 \\
1.0 & -2.0 & subgiant & 0.03 & 6824 & 853 & 20.4 & 664 & 83.0 & 2.0 \\
\hline
\multicolumn{10}{|l|}{ \textbf{Model Class 5: solar metallicity, $1.4M_\odot$} }\\\hline
1.4 & 0.0 & MS & 0.01 & 2357 & 295 & 7.0 & 229 & 28.6 & 0.7 \\
1.4 & 0.0 & subgiant & 0.03 & 8337 & 1042 & 24.9 & 811 & 101 & 2.4 \\
\hline
\multicolumn{10}{|l|}{ \textbf{Model Class 6: solar metallicity, $2.5M_\odot$} }\\\hline
2.5 & 0.0 & MS & 0.11 & 29377 & 3672 & 87.8 & 2861 & 358 & 8.6 \\
2.5 & 0.0 & subgiant & 0.18 & 48027 & 6003 & 144 & 4675 & 584 & 14.0 \\
\hline
\multicolumn{10}{|l|}{ \textbf{Model Class 7: solar metallicity, $5.0M_\odot$} }\\\hline
5.0 & 0.0 & MS & 0.05 & 13734 & 1717 & 41.1 & 1340 & 168 & 4.0 \\
5.0 & 0.0 & subgiant & 0.07 & 19636 & 2455 & 58.7 & 1915 & 239 & 5.7 \\

\hline
\end{tabular}
\caption{Same as Table~\ref{table:TESS_solar}, but for upper main-sequence pulsators (including $\delta$ Scutis).
We adopt a minimum amplitude estimate of 14~ppm for all cases.}
\label{table:TESS_dSct}
\end{table*}

%% file: tables/table8_compare_to_TESS_Mira.tex
\begin{table*}[ht]
\centering
\begin{tabular}{| c | c | c | c | c | c| c | c| c| c|} %%% 10 columns
\toprule
\multicolumn{1}{|c|}{\textbf{mass}} &
\multicolumn{1}{c|}{\textbf{[Fe/H]}} &
\multicolumn{1}{c|}{\textbf{Phase}} &
\multicolumn{1}{c|}{\textbf{$dP_\text{dmesh}$}} &
\multicolumn{1}{c|}{\textbf{ $\dPm$/$\sigma_{\text{obs}, P}$}} &
\multicolumn{1}{c|}{\textbf{$\dPm$/$\sigma_{\text{obs}, P}$}} &
\multicolumn{1}{c|}{\textbf{$\dPm$/$\sigma_{\text{obs}, P}$}} &
\multicolumn{1}{c|}{\textbf{$\dPm$/$\sigma_{\text{obs}, P}$}} &
\multicolumn{1}{c|}{\textbf{$\dPm$/$\sigma_{\text{obs}, P}$}} &
\multicolumn{1}{c|}{\textbf{$\dPm$/$\sigma_{\text{obs}, P}$}} 
\\
\hline
& 
&
& 
days &
T=4 yrs &
T=1 yrs &
T=0.083 yrs &
T=4 yrs &
T=1 yrs &
T=0.083 yrs \\ 
& 
&
& 
&
$\sigma_\text{noise}$=19.1 &
$\sigma_\text{noise}$=19.1 &
$\sigma_\text{noise}$=19.1 &
$\sigma_\text{noise}$=197 &
$\sigma_\text{noise}$=197 &
$\sigma_\text{noise}$=197 \\ 
\hline
\multicolumn{10}{|l|}{ \textbf{Model Class 1: Solar-like} }\\\hline
1.0 & 0.0 & AGB & 98.2 & 669775 & 87355 & 2095 & 65017 & 8480 & 203 \\
\hline
\multicolumn{10}{|l|}{ \textbf{Model Class 2: metal-rich, solar mass} }\\\hline
1.0 & +0.5 & AGB & 389.7 & 68532 & 8568 & 205 & 6650 & 831 & 19.9 \\
\hline
\multicolumn{10}{|l|}{ \textbf{Model Class 3: metal-poor, solar mass} }\\\hline
1.0 & -1.0 & AGB & 37.5 & 112225 & 41760 & 1442 & 10898 & 4055 & 140 \\
\hline
\multicolumn{10}{|l|}{ \textbf{Model Class 4: very metal-poor, solar mass} }\\\hline
1.0 & -2.0 & AGB & 23.7 & 1612 & 786 & 77.8 & 157 & 76.4 & 7.6 \\
\hline
\multicolumn{10}{|l|}{ \textbf{Model Class 5: solar metallicity, $1.4M_\odot$} }\\\hline
1.4 & 0.0 & AGB & 77.2 & 476293 & 65242 & 1571 & 46238 & 6334 & 152 \\
\hline
\multicolumn{10}{|l|}{ \textbf{Model Class 6: solar metallicity, $2.5M_\odot$} }\\\hline
2.5 & 0.0 & AGB & 34.1 & 380288 & 87437 & 2278 & 36925 & 8490 & 221 \\
\hline
\multicolumn{10}{|l|}{ \textbf{Model Class 7: solar metallicity, $5.0M_\odot$} }\\\hline
5.0 & 0.0 & AGB & 200.7 & 1627421 & 223467 & 5380 & 157988 & 21694 & 522 \\

\hline
\end{tabular}
\caption{Same as Table~\ref{table:TESS_solar}, but for Miras. We estimate oscillation amplitudes based on the fitted luminosity-amplitude relation (Equation~\ref{eq:amp-lum}).}
\label{table:TESS_Mira}
\end{table*}